\documentclass[aps,prx,twocolumn,superscriptaddress,print]{revtex4-1}
\usepackage[colorlinks=true,linkcolor=blue,citecolor=blue,urlcolor=blue]{hyperref}
\usepackage{amssymb,amsmath}
\usepackage{graphicx}
\usepackage{multirow}
\usepackage{colortbl}
\usepackage[table]{xcolor}
\usepackage{appendix}
\usepackage{mathtools}
\usepackage{amsmath}
\usepackage{verbatim}
\usepackage{gensymb}
\usepackage{subfigure}
\usepackage{bm}
\usepackage{threeparttable}
\usepackage{booktabs}
\usepackage{braket}
\usepackage{xcolor}
\usepackage[normalem]{ulem}
\usepackage{array, makecell}
\usepackage{pifont} 

\hypersetup{
  pdfauthor = {Yunhua Wang, Guodong Yu, Malte R\"{o}sner, Mikhail I. Katsnelson, Hai-Qing Lin, Shengjun Yuan},
  pdftitle = {Polarization-dependent selection rules and optical spectrum atlas of twisted bilayer graphene quantum dots}
}
\begin{document}
\title{Polarization-dependent selection rules and optical spectrum atlas of twisted bilayer graphene quantum dots}
\author{Yunhua Wang}
\email{wangyunhua@csrc.ac.cn}
\affiliation{Beijing Computational Science Research Center, Beijing, 100193, China}
\affiliation{Institute for Molecules and Materials, Radboud University, Heijendaalseweg 135, NL-6525 AJ Nijmegen, Netherlands}
\author{Guodong Yu}
\affiliation{Key Laboratory of Artificial Micro- and Nano-structures of Ministry of Education and School of Physics and Technology, Wuhan University, Wuhan 430072, China}
\author{Malte R\"{o}sner}
\affiliation{Institute for Molecules and Materials, Radboud University, Heijendaalseweg 135, NL-6525 AJ Nijmegen, Netherlands}
\author{Mikhail I. Katsnelson}
\affiliation{Institute for Molecules and Materials, Radboud University, Heijendaalseweg 135, NL-6525 AJ Nijmegen, Netherlands}
\author{Hai-Qing Lin}
\affiliation{Beijing Computational Science Research Center, Beijing, 100193, China}
\author{Shengjun Yuan}
\email{s.yuan@whu.edu.cn}
\affiliation{Key Laboratory of Artificial Micro- and Nano-structures of Ministry of Education and School of Physics and Technology, Wuhan University, Wuhan 430072, China}
\affiliation{Beijing Computational Science Research Center, Beijing, 100193, China}
\affiliation{Institute for Molecules and Materials, Radboud University, Heijendaalseweg 135, NL-6525 AJ Nijmegen, Netherlands}
\thanks{$^*$ Equal contribution.}
\begin{abstract}
Finding out how symmetry encodes optical polarization information into the selection rule in molecules and materials is important for their optoelectronic applications including spectroscopic analysis, display technology and quantum computation. Here, we extend the polarization-dependent selection rules from atoms to solid systems with point group descriptions via rotational operator for circular polarization and $2$-fold rotational operator (or reflection operator) for linear polarization. As a variant of graphene quantum dot (GQD), twisted bilayer graphene quantum dot (TBGQD) certainly inherits GQD's advantages including ultrathin thickness, excellent biocompatibility and shape- and size-tunable optical absorption/emission. We then naturally ask how the electronic structures and optical properties of TBGQDs rely on size, shape, twist angle and correlation effects. We build plentiful types of TBGQDs with $10$ point groups and obtain the optical selection rule database for all types, where the current operator matrix elements identify the generalized polarization-dependent selection rules. Our results show that both of the electronic and optical band gaps follow power-law scalings and the twist angle has the dominant role in modifying the size scaling. We map an atlas of optical conductivity spectra for both size and twist angle in TBGQDs. As a result of quantum confinement effect of finite size, in the atlas a new type of optical conductivity peaks absent in twisted bilayer graphene bulk is predicted theoretically with multiple discrete absorption frequencies from infrared to ultraviolet light, enabling applications on photovoltaic devices and photodetectors. The atlas and size scaling provide a full structure/symmetry-function interrelation and hence offers an excellent geometrical manipulation of optical properties of TBGQD as a building block in integrated carbon optoelectronics.
\end{abstract}

\maketitle
\section{Introduction}

Optical polarization, i.e., the oscillating direction of electric field, can be well generated, controlled and detected by means of the light-matter interaction, where various polarization-dependent phenomena of light emerge, such as birefringent, dichroism, optical activity, Kerr effect, and so forth. Besides triggering these interesting optical physics, optical polarization together with these effects has widespread applications in photodetectors\cite{goldstein2017}, laser, display technologies, spectroscopic analysis\cite{kliger2012} and quantum computation\cite{fox2006,langer2018lightwave}.
Optical selection rule specifies the possible transitions among energy states via absorption or emission of electromagnetic radiation in several physical systems from atoms to molecules and solids. The rule is essentially a strictly constrained result of both system symmetries and conservation laws, and hence is of vital importance in understanding optical spectrum and determining the symmetry and electronic states of system. It is known that, besides the parity selection rule from the angular momentum conservation, the selection rules of both circularly and linearly polarized light for hydrogen atoms in electric dipole approximation are described by the magnetic quantum number changes $\Delta m=\pm 1$ and $\Delta m=0$\cite{demtroder2010}, respectively. For the past several years, it has also been indicated that, as a result of constraints from lattice symmetry and time-reversal symmetry, optical interband transitions for left and right circular polarizations in some special semiconductors display quite different behaviors. In addition, polarization-dependent selection rules are deeply tied to the internal quantum degrees of freedom of Bloch electrons, such as the spin/valley contrasting optical selection rules for circularly polarized light in III-V bulk crystals\cite{ivchenko1978}/two-dimensional hexagonal semiconductors\cite{yao2008valley,xiao2012coupled,cao2012valley,mak2012control,zeng2012valley}. Recently, numerical calculations on the current operator matrix element, in graphene quantum dot (GQD) without Bloch bands, have shown anomalous distribution patterns with respect to the rotational symmetry operator\cite{kavousanaki2015}, as a result of the selection rule of polarized light. A natural question is whether the spin/valley contrasting optical selection rules in crystals and the anomalous optical selection rules in quantum dots can be extended to a uniform expression of polarization-dependent selection rules. In addition, understanding how symmetry encodes polarization of light into the selection rule is very meaningful for analyzing both the polarization-dependent optical spectrum and the symmetry of electronic structures in solid materials and subsequently enabling their applications in optoelectronics. Motivated by these inspirations, herein, we generalize a uniform formula of optical selection rules for both linear and circular polarizations in molecules and solids with point group descriptions. Our theory shows that instead of the magnetic quantum number change in atoms the rotational quantum number change characterizes the selection rules of circularly polarized light by virtue of the rotational operator of system. The selection rules of linearly polarized light in $D_{n}$, $D_{nh}$, $D_{nd}$ and $C_{nv}$ systems can also be correspondingly characterized by a $2$-fold rotational operator or reflection operator of systems.

Compared with bulk semiconductors, quantum dots, i.e., zero-dimensional nanocrystals, possess remarkably size- and shape-tunable energy levels and absorption/emission spectra because of quantum confinement effects\cite{efros1982,brus1984electron}, and subsequently enable a wide range of optoelectronic devices\cite{kagan2016building,won2019highly,kim2020efficient,liu2020micro}, such as displays, solar cells and light-emitting diodes. GQD (i.e., graphene nanofragments) has ultrathin thickness, excellent biocompatibility, easy functionalization, good photostability, and shape- and size-controllable optical absorptions as well as photoluminescence, and hence brings promising applications on optical sensors\cite{shen2012graphene,li2013focusing}, bioimaging\cite{zheng2015glowing,shen2012graphene,li2013focusing}, photovoltaics\cite{zhang2012graphene,bacon2014graphene,yan2019recent}, photodetectors\cite{zhu2015photoluminescence} and light-emitting diodes\cite{zhang2012graphene,bacon2014graphene,yan2019recent}. Several synthetic strategies are used to fabricate successfully GQDs from a few nm to several hundred nm\cite{ponomarenko2008chaotic,ritter2009influence,kim2012anomalous,shen2012graphene,zhang2012graphene,
li2013focusing,bacon2014graphene,bacon2014graphene,
zheng2015glowing,zhu2015photoluminescence,yan2019recent}. Theoretical investigations based on tight-binding model reveal well that the optical absorption of GQDs is modulated by the edge type\cite{yamamoto2006edge}, size\cite{zhang2008tuning,basak2015theory}, shape\cite{pohle2018symmetry} and electron-correlation effects\cite{ozfidan2014microscopic,basak2015theory}. In addition, the group theory analysis shows that the symmetry of GQDs plays a key role in optical selection rules\cite{pohle2018symmetry}. Recently, twisted bilayer graphene has drawn considerable attention in condensed matters owing to their exotic electronic structure\cite{dos2007graphene,morell2010flat,trambly2010localization,bistritzer2011moire}, emergent correlated effects\cite{cao2018unconventional,bernevig2021twisted} and quasicrystalline order\cite{ahn2018dirac,yao2018quasicrystalline} in these systems. The optical absorption properties of infinite-size twisted bilayer graphene and has been explored theoretically\cite{moon2013optical,le2018electronic,vela2018electronic,yu2019dodecagonal}, based on the $p_z$ orbital based tight-binding model. It is natural to further ask how the electronic structures and optical properties rely on size, shape, twist angle, edge structure and correlation effects in twisted bilayer graphene quantum dots (TBGQDs). The chiral optical properties including optical activity and circular dichroism are analyzed in TBGQDs with a special $D_n$ point group symmetry (with $n=2,3,6$)\cite{tepliakov2020twisted}. Here, via a combination among the twist angle, geometrical center and edge restrictions we have searched various TBGQD structures with $10$ different point groups. Applying the orthogonality theorem we evaluate the optical selection rules for all these structures. The calculated results on current operator matrix elements follow indeed the generalized polarization-dependent selection rules. The band gap of these quantum dots follows power-law scalings with their power indexes inside $[-2,-1]$. As a consequence of selection rules, optical conductivity spectrum exhibit three remarkable absorption characteristics: (i) a relatively strong absorption occur at about $1.5t_0<\hbar\omega<2t_0$ (with $t_0=2.8$ eV), which is associated to the interband transitions between these energy levels near the Van Hove singularities of twisted bilayer graphene; (ii) the optical band gap scaling also follows the power-law but with its power index less than that of electronic band gap due to the possible forbidden transitions between valence and conduction band edges; and (iii) the quantum confinement effect of finite size renders a new type of optical conductivity peaks besides the previous three types of conductivity peaks of infinite twisted bilayer graphene, and the new group of peaks with multiple discrete absorption frequencies from infrared to ultraviolet light enables potential applications on photovoltaic devices and photodetectors.

\begin{table*}[htbp!]
\caption{{Summary on irreducible representations for $z$, $x$ and $y$ in point group character tables.} The \checkmark (\ding{55}) sign denotes the distinguishable (indistinguishable) irreducible representations between $z$ in the second row and $(x,y)$ in the third row and between $x$ and $y$ inside $(x,y)$. $\hat{C}_n(z)$ is the rotational symmetry operator, i.e., $\hat{R}_{2\pi/n}$. The symmetry operator $\hat{O}$ for linearly polarized light can be $\hat{C}_2$, $\hat{C}^{\prime(\prime\prime)}_2$ or $\hat{\sigma}_{v(d)}$, with corresponding $n$ inside $\{\}$. For instance, $\{3, 5\}:\hat{C}_2/\hat{\sigma}_v$ for $D_{nh}$ represents $\hat{O}=\hat{C}_2$ or $\hat{\sigma}_v$ for $D_{3h}$ and $D_{5h}$ point groups.}
\centering
\begin{tabular}{c c c c c c c c c c c c}
\hline\hline
 & Nonaxial & $C_n$ & $D_{n}$ & $C_{nv}$ & $C_{nh}$ & $D_{nh}$ & $D_{nd}$ & $S_n$ & Cubic & $C_{\infty v}$ & $D_{\infty h}$ \\ \hline

 $z$ & \makecell{$C_s$(\checkmark)\\ $C_{1(i)}$(\ding{55})}  & \checkmark &  \checkmark  & \checkmark & \checkmark & \checkmark & \checkmark & \checkmark & \ding{55}& \checkmark & \checkmark \\
 $(x,y)$ &  & \ding{55} & \makecell{$D_2$(\checkmark)\\ others(\ding{55})} & \makecell{$C_{2v}$(\checkmark)\\ others(\ding{55})} & \ding{55} & \makecell{$D_{2h}$(\checkmark)\\ others(\ding{55})}  & \ding{55} & \ding{55} & \ding{55} & \ding{55} & \ding{55}\\
 $\hat{C}_n(z)$  &  &  $\hat{R}_{2\pi/n}$ &  $\hat{R}_{2\pi/n}$ &  $\hat{R}_{2\pi/n}$ &  $\hat{R}_{2\pi/n}$ & $\hat{R}_{2\pi/n}$ &  $\hat{R}_{2\pi/n}$ &  $\hat{R}_{2\pi/n}$ &  &  $\hat{R}_{2\pi/n}$ & $\hat{R}_{2\pi/n}$ \\
 $\hat{O}$ &  &  & \makecell{$\{3, 5\}:\hat{C}_2$ \\ $\{4, 6\}:\hat{C}^{\prime(\prime\prime)}_2$} & \makecell{$\{3, 5\}:\hat{\sigma}_v$ \\ $\{4, 6\}:\hat{\sigma}_{v(d)}$} &   & \makecell{$\{3, 5\}:\hat{C}_2/\hat{\sigma}_v$ \\ $\{4, 6,8\}:\hat{C}^{\prime(\prime\prime)}_2/\hat{\sigma}_{v(d)}$} & \makecell{$\{3, 5\}:\hat{C}_2/\hat{\sigma}_d$ \\ $\{2, 4, 6\}:\hat{C}^\prime_2/\hat{\sigma}_d$} &  &   & $\infty \hat{\sigma}_v$ & \makecell{$\infty \hat{\sigma}_v$\\$\infty \hat{C}_2$}\\
  \hline \hline
  \end{tabular}
 \label{table:pg}
 \end{table*}

\section{Polarization-dependent selection rules}

The optical conductivity formula indicates that the allowed transitions are determined by the nonzero matrix elements of the current operator $\hat{j}_\alpha$ with $\alpha = x, y, z$, i.e.,
\begin{equation}
\braket {\psi_k |\hat{j}_\alpha |\psi_l}\neq 0,
\label{eq:selrule}
\end{equation}
where $|\psi_k \rangle$ and $|\psi_l \rangle$ are two eigenstates of system. It is also a fact that these nonzero matrix elements can be picked up by virtue of the orthogonality theorem in group theory. The reducible representation of $\hat{j}_\alpha |\psi_l\rangle$ is the direct product representation, $\Gamma_{\hat{j}_\alpha}\otimes\Gamma_l$, which is usually written as a direct sum form $\sum^\oplus_\mu a_\mu\Gamma_\mu$ with $a_\mu$ as the number of times the irreducible representation $\Gamma_\mu$ appears.
If the current operators $\hat{j}_x$, $\hat{j}_y$ and $\hat{j}_z$ have different irreducible representations, the selection rules for different linearly polarized light are naturally distinguishable. From character tables of point groups\cite{cotton2003chemical,dresselhaus2007group}, we can conclude that, (i) except $C_i$ and cubic point groups, all other point groups have different irreducible representations for $z$ and $(x, y)$, and (ii) except the nonaxial groups, $D_2$, $C_{2v}$ and $D_{2h}$ point groups, all the other point groups have the same irreducible representations for $x$ and $y$, as listed in Table~\ref{table:pg}. The irreducible representation of $\hat{j}_\alpha$ is the same as that of $\alpha$. In addition, considering that many of 2D materials and their quantum dot structures have non-cubic point group symmetry, we thus mainly seek theoretical formula governing the selection rules of polarized light for these point groups with the same irreducible representations of $\hat{j}_x$ and $\hat{j}_y$. In this respect, we need to search for some symmetry operators to differentiate $\Gamma_{\hat{j}_{x(y)}}\otimes\Gamma_l$ for linear polarization and $\Gamma_{\hat{j}_\pm}\otimes\Gamma_l$ for right $(\sigma^+)$ and left $(\sigma^-)$ circular polarizations with $\hat{j}_\pm=\hat{j}_x\pm i\hat{j}_y$.

For circularly polarized light, we choose the rotational symmetry operator $\hat{C}_n (\hat{R}_{2\pi/n})$ with a $n$-fold $z$ axis, where an arbitrary eigenstate $|\psi_l \rangle$ of system with the same $\Gamma_l$ is distinguished by the eigenvalues (i.e., rotational quantum number $\phi_l$) of $\hat{C}_n$, as follows:
\begin{equation}
\hat{C}_n|\psi_l \rangle=e^{\frac{i2\pi}{n}\phi_l}|\psi_l \rangle,
\end{equation}
with $\phi_l=0,\cdots,n-1$. For an allowed transition $k\leftrightarrow l$, the matrix elements of current operator $\hat{j}_\pm$ for circular polarizations take the forms as
\begin{equation}
\braket {\psi_k |\hat{j}_\pm |\psi_l}=e^{\frac{i2\pi}{n}(\phi_k-\phi_l)}\braket {\psi_k |\hat{C}^\dagger_n\hat{j}_\pm \hat{C}_n|\psi_l}.
\label{eq:cmatele}
\end{equation}
In addition, $\hat{j}_\pm $ under the rotational transformation take the forms as
\begin{equation}
\hat{C}^\dagger_n\hat{j}_\pm \hat{C}_n = e^{\pm\frac{i2\pi}{n}}\hat{j}_\pm.
\label{eq:transf}
\end{equation}
Substituting Eq.~\eqref{eq:transf} into Eq.~\eqref{eq:cmatele} and using  Eq.~\eqref{eq:selrule}, we obtain $\phi_k-\phi_l\pm 1 =nm$ with an integer $m$. However, the constraint of $|\phi_k-\phi_l|\leq n-1$ for both left and right circular polarizations enables $m=0$. Therefore, the selection rules of circularly polarized light read
\begin{equation}
\bigtriangleup\phi= \phi_l-\phi_k=\pm 1.
\label{eq:cselrule}
\end{equation}
Eq.~\eqref{eq:cselrule} indicates that the rotational quantum number change characterizes the absorption and emission of circularly polarized light in these systems with rotational symmetry operator $\hat{C}_n$.

For linearly polarized light, we also need to find a symmetry operator $\hat{O}$ to characterize its selection rule. The operator $\hat{O}$ should satisfy two conditions: (i) $[\hat{H}, \hat{O}]=0$ enables $\hat{O}|\psi_i \rangle=\gamma_i|\psi_i \rangle$ with its eigenvalue $\gamma_i$, and (ii) $\hat{j}_\alpha$ under the transformation of $\hat{O}$ has a form as $\hat{O}^\dagger\hat{j}_\alpha\hat{O} = q_\alpha\hat{j}_\alpha$ with $q_x\neq q_y$ to distinguish $\hat{j}_x$ and $\hat{j}_y$. For an allowed transition $k\leftrightarrow l$, the matrix elements of current operator $\hat{j}_\alpha$ for linear polarizations read
\begin{equation}
\braket {\psi_k |\hat{j}_\alpha |\psi_l}=q_\alpha(\gamma^*_k)^{-1}(\gamma_l)^{-1}\braket {\psi_k |\hat{j}_\alpha|\psi_l}.
\label{eq:lmatele}
\end{equation}
Substituting Eq.~\eqref{eq:selrule} into Eq.~\eqref{eq:lmatele}, we obtain
\begin{equation}
\gamma^*_k \gamma_l - q_\alpha =0.
\label{eq:lselrule}
\end{equation}
Eq.~\eqref{eq:lselrule} characterizes the absorption and emission of linearly polarized light under the basis functions of $\hat{O}$. The existing symmetry operators $\hat{O}$ for $D_n$, $C_{nv}$, $D_{nh}$, $D_{nd}$, $C_{\infty v}$ and $D_{\infty h}$ point groups are listed in Table~\ref{table:pg}, where $\hat{O}$ is absent for $C_n$, $C_{nh}$ and $S_{n}$ point groups.

\begin{figure}[!htbp]
\centering
\includegraphics[width=8.5 cm]{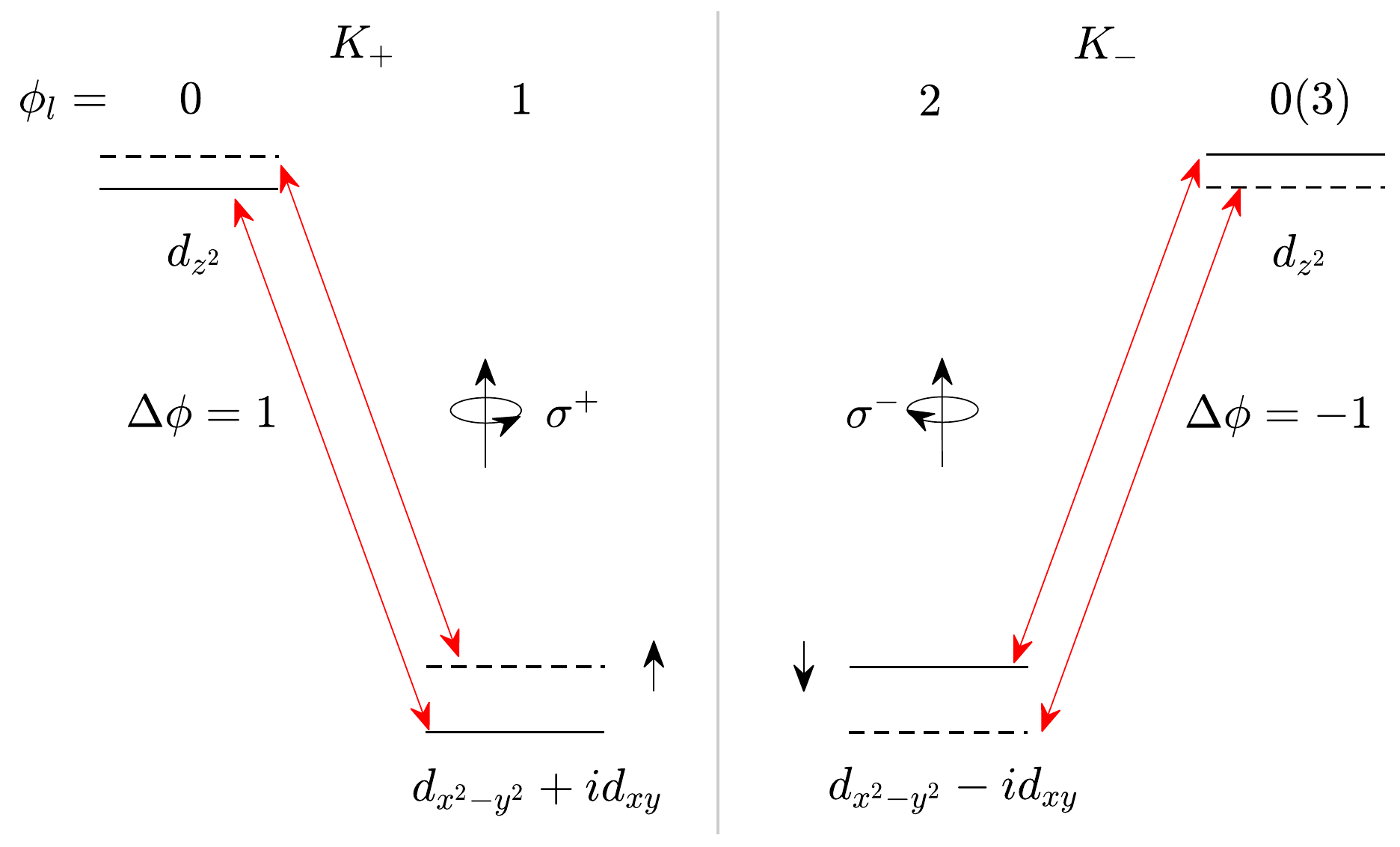}
\caption{{The selection rule of circularly polarized light in MoS$_2$ monolayer.} The conduction band bottom is mainly contributed by $d_{z^2}$ orbital of Mo, and the top of valence band edges are mainly contributed by $\{d_{x^2-y^2}, d_{xy}\}$ orbitals of Mo. These band edges with spin up (dash lines) and spin down (solid lines) are classified by the rotational quantum number $\phi_l$ of $\hat{C}_3$ at $K_+$ (left panel) and $K_-$ (right panel) valleys.}
\label{fig:cMoS2}
\end{figure}

\begin{figure*}[!htbp]
\centering
\includegraphics[width=17 cm]{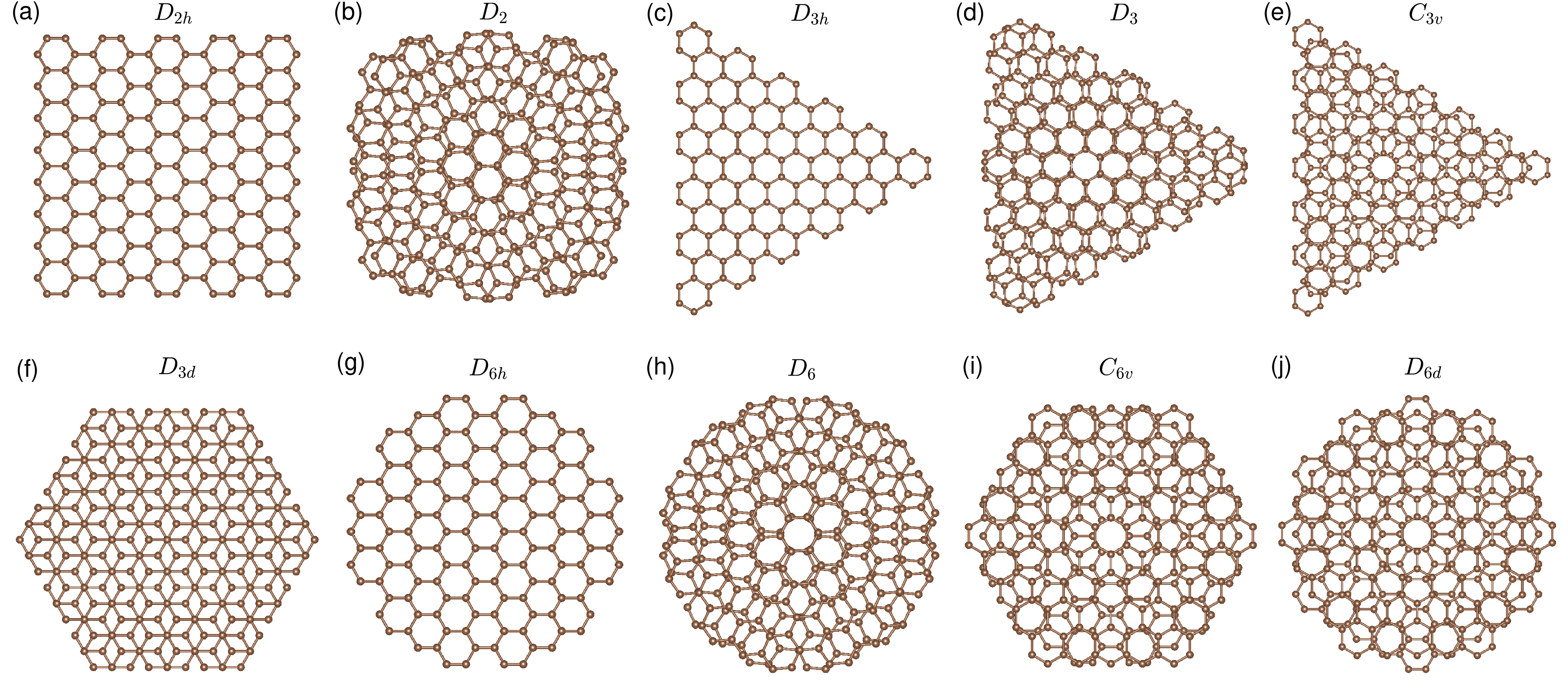}
\caption{{Top views of TBGQDs with $10$ different point group symmetries.} The structures with $D_{nh}$ have zero twist angle and hence the top layer covers vertically the bottom layer from top view in (a), (c) and (g) owing to the horizontal reflection $\sigma_h$. The structure with $D_{3d}$ in (f) has the AB stacking configuration. In the sketched $D_n$ structures the twist angles are chosen as $10^\circ$ for $D_2$ in (b), $5^\circ$ for $D_3$ in (d), and $10^\circ$ for $D_6$ in (h), respectively. These structures with $C_{3v}$ in (e), $C_{6v}$ in (i) and $D_{6d}$ in (j) have the fixed twist angles of $30^\circ$. All the symmetry operator elements and $x$ axis for these $10$ point groups are listed in supplemental Table S1\cite{SM}.}
\label{fig:structures}
\end{figure*}

Now we turn back to examine the polarization-dependent selection rules happening in the two previously studied systems including GQDs\cite{kavousanaki2015,pohle2018symmetry} and two-dimensional hexagonal semiconductors\cite{yao2008valley,xiao2012coupled,cao2012valley,mak2012control,zeng2012valley}. As an example, we consider MoS$_2$ monolayers with the $D_{3h}$ point group symmetry, where its conduction and valence band edges are mainly contributed by $d_{z^2}$ and $\{d_{x^2-y^2}, d_{xy}\}$ orbitals of Mo atoms, respectively. The rotational symmetry operator $\hat{C}_3 (\hat{R}_{2\pi/3})$ in the bases of $[d_{x^2-y^2}, d_{xy}, d_{z^2}]$ has the following representation,
\begin{equation}
\begin{split}
\hat{C}_3 [d_{x^2-y^2}, d_{xy}, d_{z^2}] = [d_{x^2-y^2}, d_{xy}, d_{z^2}]
\begin{bmatrix}
-\frac{1}{2} & \frac{\sqrt{3}}{2} & 0 \\
-\frac{\sqrt{3}}{2} & -\frac{1}{2} & 0 \\
0 & 0 & 1\\
\end{bmatrix}.\\
\end{split}
\label{eq:MoS2_C3}
\end{equation}
Using Eq.~\eqref{eq:MoS2_C3}, we can classify band edges by the rotational quantum number $\phi_l$, i.e.,
\begin{equation}
\begin{split}
d_{\phi_l=0} &= d_{z^2},\\
d_{\phi_l=1} &= \frac{1}{\sqrt{2}}(d_{x^2-y^2}+i d_{xy}),\\
d_{\phi_l=2} &= \frac{1}{\sqrt{2}}(d_{x^2-y^2}-i d_{xy}).\\
\end{split}
\end{equation}
On the other hand, we should also consider the constraint $E_{n,K_+,s}\neq E_{n,K_-,s}$ with energy $E$ from the broken inversion symmetry and the constraint $E_{n,K_+,s}= E_{n,K_-,-s}$ from the time-reversal symmetry with spin $s$ and two valley indexes $K_+$ and $K_-$. Therefore, $d_{\phi_l=1}$ and $d_{\phi_l=2}$ orbitals related by the time-reversal operation should locate at $K_+$ and $K_-$, respectively, as shown in Fig. \ref{fig:cMoS2}. The selection rule of circularly polarized light, $\bigtriangleup\phi=\pm 1$ in Eq.~\eqref{eq:cselrule}, reflects the valley-contrasted optical absorption from the analysis of azimuthal quantum number change of Mo atoms\cite{cao2012valley,mak2012control,zeng2012valley}. For GQDs with $C_{3v}$ and $C_{6v}$ point group symmetries, our selection rules in Eq.~\eqref{eq:cselrule} directly enable the transitions from $\phi_l$ to $\phi_l-1$ and $\phi_l+1$ for the corresponding $\sigma^+$ and $\sigma^-$ circularly polarized light, and hence agree with well the previous numerical calculations\cite{kavousanaki2015,pohle2018symmetry}.

\begin{figure}[!htbp]
\centering
\includegraphics[width=8.5 cm]{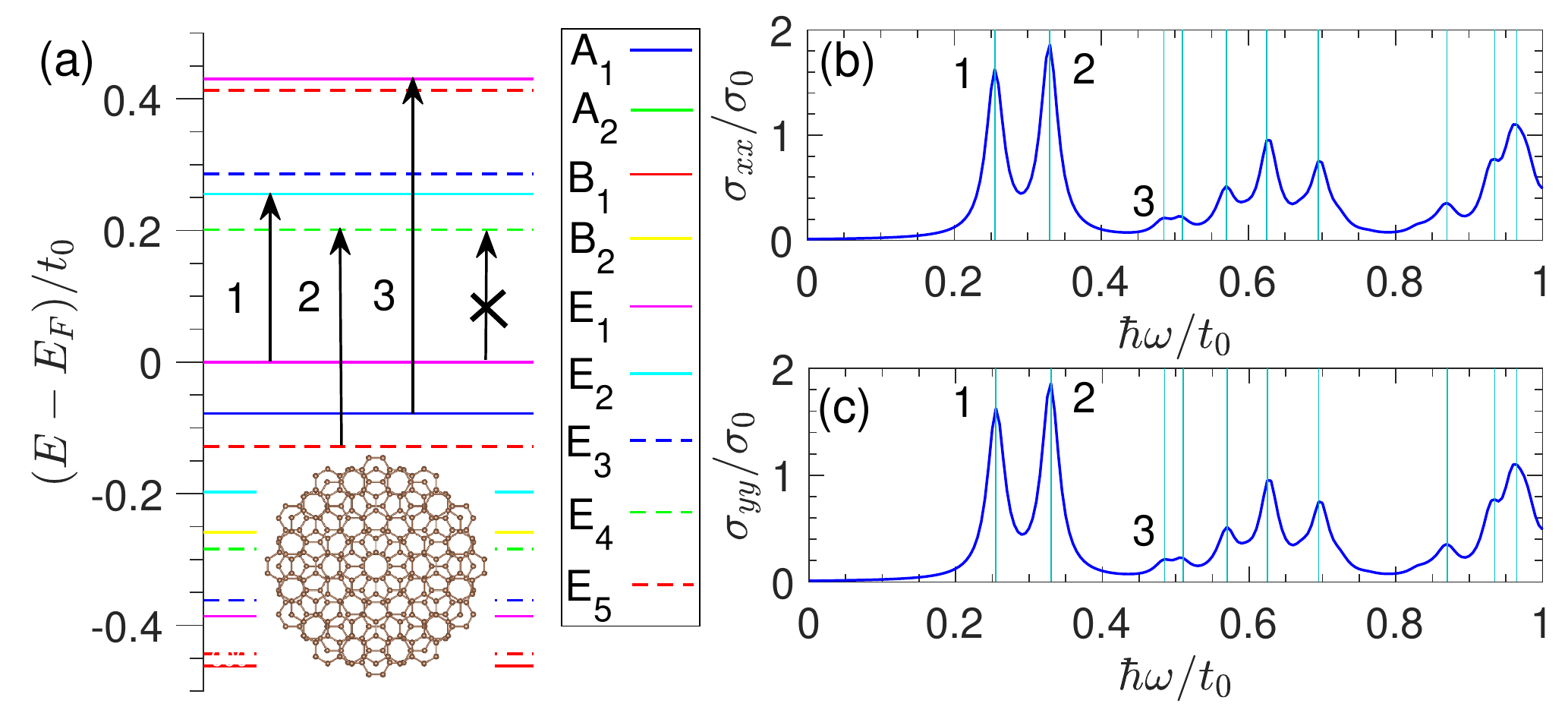}
\caption{{Energy spectrum and optical conductivity spectrum of a $D_{6d}$ TBGQD.} (a) The energy spectrum with its irreducible representation. The Fermi energy has been shifted to $0$ eV, and the inset shows the structure with $N=300$. The real part of optical conductivity as a function of photon energy $\hbar\omega$: $\sigma_{xx}$ in (b) and  $\sigma_{yy}$ in (c) in unit of $\sigma_0 = \pi e^2/(4\hbar)$. The peaks $1$, $2$ and $3$ in (b) and (c) arise from the corresponding transitions in (a). The transition $E_1 \leftrightarrow E_4$ is forbidden as a result of the selection rule.}
\label{fig:OPs_D6d}
\end{figure}

\begin{figure*}[!htbp]
\centering
\includegraphics[width=17 cm]{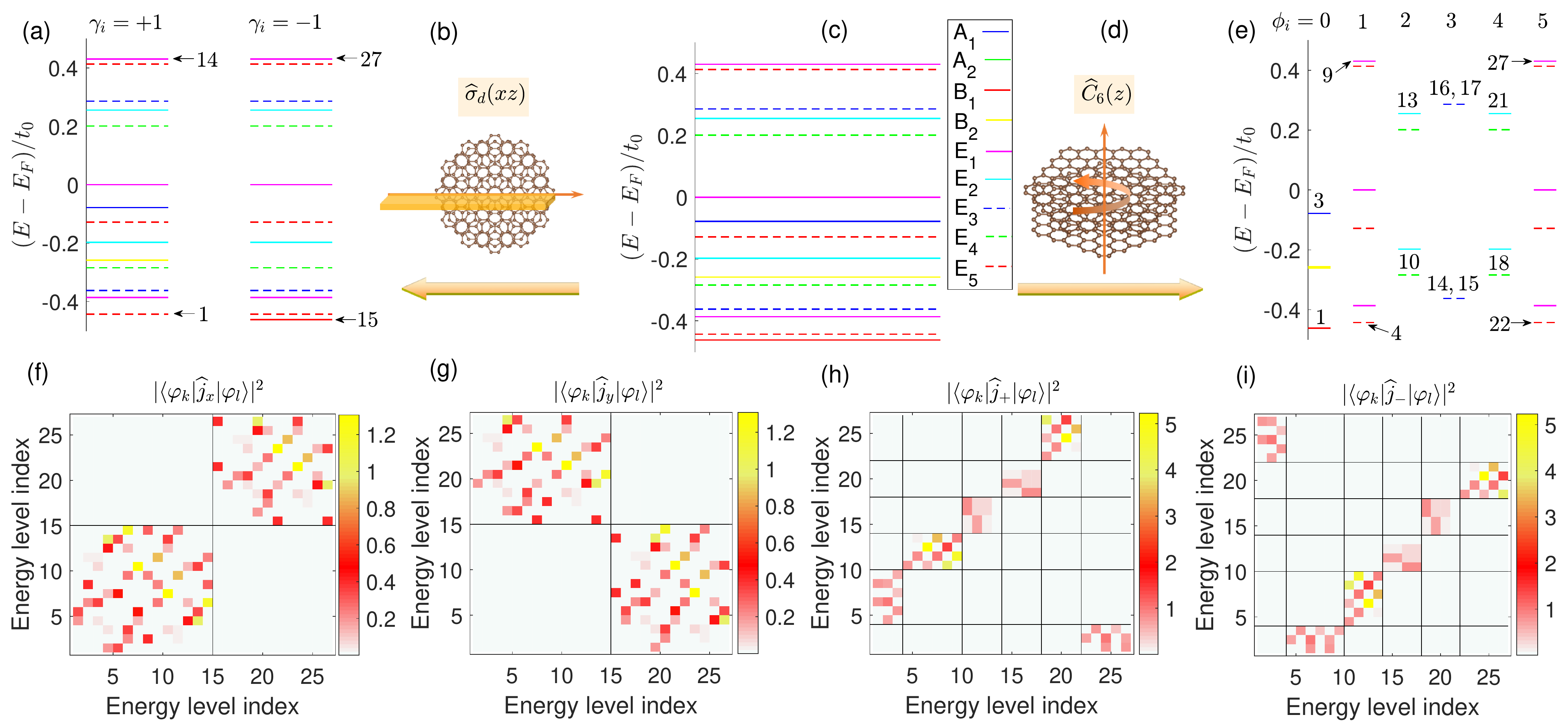}
\caption{{The classified energy spectra and the matrix elements of current operators for a $D_{6d}$ TBGQD under $\hat{\sigma}_d(xz)$ and $\hat{C}_6(z)$ operations.} The pristine energy spectrum of the $D_{6d}$ structure with $N=300$ is plotted in (c). (a,e) The classified energy spectra by $\gamma_i$ of $\hat{\sigma}_d(xz)$ in (b) with a vertical (yellow) reflection plane $xz$ and by $\phi_i$ of $\hat{C}_6$ in (d) with respect to the $z$ axis, respectively. The big yellow arrows in (b) and (d) only signal the classified results in (a) and (e) under corresponding operations. These energy levels from down to up are denoted by the increasing numbers for each $\gamma_l$ and $\phi_l$. The square modulus of current operator matrix elements in unit of $t_0^2$: $|\braket {\psi_k |\hat{j}_x |\psi_l}|^2$ in (f) and $|\braket {\psi_k |\hat{j}_y|\psi_l}|^2$ in (g) under operator $\hat{\sigma}_d(xz)$ for linearly polarized light, and $|\braket {\psi_k |\hat{j}_+ |\psi_l}|^2$ in (h) and $\braket {|\psi_k |\hat{j}_-|\psi_l}|^2$ in (i) under operator $\hat{C}_6$ for circularly polarized light. The energy level index in (f)-(i) represents these energy levels with corresponding numbers in (a) and (e).}
\label{fig:sel_D6d}
\end{figure*}

\begin{table*}
\caption{{Direct product $\Gamma_{j_\alpha}\otimes\Gamma_l$, selection rules and symmetry operator $\hat{O}$ for TBGQDs.}  The $10$ different point groups are the same as those in Fig. \ref{fig:structures}. For $D_2$ and $D_{2h}$, the selection rules for $x$ and $y$ polarized light are intrinsically distinguishable owing to the different irreducible representations of $x$ and $y$. Here, $\hat{O}^\dagger\hat{j}_\alpha\hat{O} = q_\alpha\hat{j}_\alpha$ with $\alpha=x,y$. The operator $\hat{O}$ has been chosen such that $q_x=1$ and $q_y=-1$ for all $10$ point groups.}
\label{table:sel}
\begin{ruledtabular}
\begin{centering}
\begin{tabular}{ccccccc}
 Point groups & $\Gamma_{j_\alpha}\otimes\Gamma_l$ & Selection rules & $\hat{O}$ & $q_x$ & $q_y$\\
\hline
$D_3$ & \makecell[c]{ $E\otimes A_1=E, E\otimes A_2=E$ \\ $E\otimes E=A_1\oplus A_2 \oplus E$}& $A_1\leftrightarrow E, A_2\leftrightarrow E, E\leftrightarrow E$ & $\hat{C}_2(x)$ & $+1$ & $-1$ \\
$D_{3h}$ & {{\colorbox{gray!15}{\thead{\makecell[c]{ $E^\prime\otimes A_1^\prime=E^\prime, E^\prime\otimes A_2^\prime=E^\prime$ \\ $E^\prime\otimes A_1^{\prime\prime}=E^{\prime\prime}, E^\prime\otimes A_2^{\prime\prime}=E^{\prime\prime}$ \\ $E^\prime\otimes E^\prime=A_1^\prime\oplus A_2^\prime \oplus E^\prime$ \\ $E^{\prime\prime}\otimes E^{\prime\prime}=A_1^{\prime\prime}\oplus A_2^{\prime\prime} \oplus E^{\prime\prime}$}}}}} &
\makecell[c]{$A_1^\prime\leftrightarrow E^\prime, A_2^\prime\leftrightarrow E^\prime, E^\prime\leftrightarrow E^\prime$ \\ $A_1^{\prime\prime}\leftrightarrow E^{\prime\prime}, A_2^{\prime\prime}\leftrightarrow E^{\prime\prime}, E^{\prime\prime}\leftrightarrow E^{\prime\prime}$} &$\hat{\sigma}_v(xz)$ & $+1$ & $-1$ \\

$C_{3v}$ &\makecell[c]{ $E\otimes A_1=E, E\otimes A_2=E$ \\ $E\otimes E=A_1\oplus A_2 \oplus E$} & $A_1\leftrightarrow E, A_2\leftrightarrow E, E\leftrightarrow E$ & $\hat{\sigma}_v(xz)$ & $+1$ & $-1$ \\

$D_{3d}$ &{{\colorbox{gray!15}{\thead{\makecell[c]{ $E_u\otimes A_{1g}=E_u, E_u\otimes A_{2g}=E_u$ \\ $E_u\otimes A_{1u}=E_g, E_u\otimes A_{2u}=E_g$  \\ $E_u\otimes E_g=A_{1u}\oplus A_{2u} \oplus E_u$ \\ $E_u\otimes E_u=A_{1g}\oplus A_{2g} \oplus E_g$}}}}} &\makecell[c]{$A_{1g}\leftrightarrow E_u, A_{2g}\leftrightarrow E_u$\\$A_{1u}\leftrightarrow E_g, A_{2u}\leftrightarrow E_g, E_g\leftrightarrow E_u$ }& $\hat{\sigma}_d(xz)$ & $+1$ & $-1$\\

$D_6$ &\makecell[c]{ $E_1\otimes A_1=E_1, E_1\otimes A_2 =E_1$ \\ $E_1\otimes B_1=E_2, E_1\otimes B_2=E_2$ \\ $E_1\otimes E_1=A_1\oplus A_2 \oplus E_2$ \\ $E_1\otimes E_2=B_1\oplus B_2 \oplus E_1$} &\makecell[c]{$A_1\leftrightarrow E_1, A_2\leftrightarrow E_1$\\$B_1\leftrightarrow E_2, B_2\leftrightarrow E_2, E_1\leftrightarrow E_2$ }& $\hat{C}_2^\prime(x)$ & $+1$ & $-1$  \\

$D_{6h}$ &{{\colorbox{gray!15}{\thead{\makecell[c]{ $E_{1u}\otimes A_{1g}=E_{1u}, E_{1u}\otimes A_{2g}=E_{1u}$ \\ $E_{1u}\otimes B_{1g}=E_{2u}, E_{1u}\otimes B_{2g}=E_{2u}$ \\
$E_{1u}\otimes A_{1u}=E_{1g}, E_{1u}\otimes A_{2u}=E_{1g}$ \\ $E_{1u}\otimes B_{1u}=E_{2g}, E_{1u}\otimes B_{2u}=E_{2g}$\\
$E_{1u}\otimes E_{1g}=A_{1u}\oplus A_{2u} \oplus E_{2u}$ \\ $E_{1u}\otimes E_{2g}=B_{1u}\oplus B_{2u} \oplus E_{1u}$ \\
$E_{1u}\otimes E_{1u}=A_{1g}\oplus A_{2g} \oplus E_{2g}$ \\ $E_{1u}\otimes E_{2u}=B_{1g}\oplus B_{2g} \oplus E_{1g}$ }}}}}
&\makecell[c]{$A_{1g}\leftrightarrow E_{1u}, A_{2g}\leftrightarrow E_{1u}$ \\$B_{1g}\leftrightarrow E_{2u}, B_{2g}\leftrightarrow E_{2u}$ \\$A_{1u}\leftrightarrow E_{1g}, A_{2u}\leftrightarrow E_{1g}$
\\$B_{1u}\leftrightarrow E_{2g}, B_{2u}\leftrightarrow E_{2g}$
\\$E_{1g}\leftrightarrow E_{2u}, E_{2g}\leftrightarrow E_{1u}$ }& $\hat{C}_2^\prime(x)$ & $+1$ & $-1$ \\

$C_{6v}$ &\makecell[c]{ $E_1\otimes A_1=E_1, E_1\otimes A_2 =E_1$ \\ $E_1\otimes B_1=E_2, E_1\otimes B_2=E_2$ \\ $E_1\otimes E_1=A_1\oplus A_2 \oplus E_2$ \\ $E_1\otimes E_2=B_1\oplus B_2 \oplus E_1$} &\makecell[c]{$A_1\leftrightarrow E_1, A_2\leftrightarrow E_1$\\$B_1\leftrightarrow E_2, B_2\leftrightarrow E_2, E_1\leftrightarrow E_2$ }& $\hat{\sigma}_v(xz)$ & $+1$ & $-1$   \\

$D_{6d}$ & {{\colorbox{gray!15}{\thead{\makecell[c]{ $E_1\otimes A_1=E_1, E_1\otimes A_2 =E_1$ \\ $E_1\otimes B_1=E_5, E_1\otimes B_2=E_5$ \\ $E_1\otimes E_1=A_1\oplus A_2 \oplus E_2$ \\
 $E_1\otimes E_2=E_1\oplus E_3, E_1\otimes E_3=E_2\oplus E_4$ \\ $E_1\otimes E_4=E_3\oplus E_5$ \\ $E_1\otimes E_5=B_1\oplus B_2 \oplus E_4$}}}}} & \makecell[c]{$A_1\leftrightarrow E_1, A_2\leftrightarrow E_1$ \\ $B_1\leftrightarrow E_5, B_2\leftrightarrow E_5$ \\ $E_1\leftrightarrow E_2, E_2\leftrightarrow E_3 $ \\ $E_3\leftrightarrow E_4, E_4\leftrightarrow E_5$ }& $\hat{\sigma}_d(xz)$ & $+1$ & $-1$  \\

$D_2$ & \makecell[c]{$B_3\otimes A=B_3, B_3\otimes B_1 =B_2$ \\ $B_3\otimes B_2 =B_1, B_3\otimes B_3 =A$ \\ $B_2\otimes A=B_2, B_2\otimes B_1 =B_3$ \\ $B_2\otimes B_2 =A, B_2\otimes B_3 =B_1$}
& \makecell[c]{$x:$ \\ $A\leftrightarrow B_3, B_1 \leftrightarrow B_2$ \\
$y:$\\ $A\leftrightarrow B_2, B_1 \leftrightarrow B_3$} & $\hat{C}_2(x)$ & $+1$ & $-1$ \\

$D_{2h}$ &{{\colorbox{gray!15}{\thead{\makecell[c]{$B_{3u}\otimes A_g=B_{3u}, B_{3u}\otimes B_{1g} =B_{2u}$ \\ $B_{3u}\otimes B_{2g} =B_{1u}, B_{3u}\otimes B_{3g} =A_{u}$ \\$B_{3u}\otimes A_u=B_{3g}, B_{3u}\otimes B_{1u} =B_{2g}$ \\ $B_{3u}\otimes B_{2u} =B_{1g}, B_{3u}\otimes B_{3u} =A_{g}$ \\
$B_{2u}\otimes A_g=B_{2u}, B_{2u}\otimes B_{1g} =B_{3u}$ \\ $B_{2u}\otimes B_{2g} =A_{u}, B_{2u}\otimes B_{3g} =B_{1u}$ \\$B_{2u}\otimes A_u=B_{2g}, B_{2u}\otimes B_{1u} =B_{3g}$ \\ $B_{2u}\otimes B_{2u} =A_{g}, B_{2u}\otimes B_{3u} =B_{1g}$}}}}}
& \makecell[c]{$x:$ \\ $A_g\leftrightarrow B_{3u}, B_{1g} \leftrightarrow B_{2u}$ \\
$B_{2g}\leftrightarrow B_{1u}, B_{3g} \leftrightarrow A_u$ \\
$y:$ \\ $A_g\leftrightarrow B_{2u}, B_{1g} \leftrightarrow B_{3u}$ \\
$B_{2g}\leftrightarrow A_u, B_{3g} \leftrightarrow B_{1u}$} & $\hat{C}_2(x)$ & $+1$ & $-1$ \\

\end{tabular}
\par\end{centering}
\end{ruledtabular}
\end{table*}

\section{Optical spectrum and matrix elements of current operators}

A combination among the twist degree of freedom, geometrical center position and edge restrictions results in a lot of twisting graphene quantum dot structures with various symmetries. Many of them can be summarized into $10$ point groups including $D_{2h}$, $D_2$, $D_{3h}$, $D_3$, $C_{3v}$, $D_{3d}$, $D_{6h}$, $D_6$, $C_{6v}$ and $D_{6d}$, as shown in Fig. \ref{fig:structures}. Combining the procedures of generating these structures (see Appendix~\ref{sec:pgQDs}), their features can be summarized as follows: (i) these structures with $D_{nh}$ point group are actually the AA stacking bilayer GQDs with a typical horizontal mirror plane, and the structure with $D_{3d}$ is the AB stacking bilayer GQDs; (ii) the geometrical centers are here at the hexagon center except $D_2$, $D_{2h}$ and $D_{3d}$, where the centers for $D_2$ and $D_{2h}$ are at the middle of bond and at atom for $D_{3d}$, respectively; (iii) the twist angles for $C_{3v}$, $C_{6v}$ and $D_{6d}$ are $\pi/6$, and the twist angles for $D_3$ and $D_6$ can change from 0 to $\pi/6$ if the geometrical center is at the hexagon center and from 0 to $ \pi/3$  if the geometrical center is at atom, because of the twist periodicity; (iv) the $x$ axes for $D_{2h}$, $D_{3h}$ and $D_{6h}$ can be fixed along zigzag or armchair directions, and the $x$ axes are chosen along $C_2(x)$ for $D_2$ and $D_3$, $\sigma_v(xz)$ for $C_{3v}$, $D_{3h}$ and $C_{6v}$, $\sigma_d(xz)$ for $D_{3d}$ and $D_{6d}$, and $C_2^\prime(x)$ for $D_6$ and $D_{6h}$. All irreducible representations, symmetry operators and $x$ axis of TBGQDs for each point group are listed in supplemental Table SI\cite{SM}.

We now apply our theory to the optical spectrum of TBGQDs. The direct product $\Gamma_{\hat{j}_\alpha}\otimes\Gamma_l$ for each irreducible representation $\Gamma_l$ and the selection rules of $10$ point groups are calculated and listed in Table~\ref{table:sel}. We firstly check the selection rules by virtue of the optical conductivity, whose real part corresponds to the interband absorption. These absorption peaks encode the information of the allowed transitions and hence manifest the selection rules. As an example, Fig. \ref{fig:OPs_D6d} shows the energy spectrum and optical conductivity spectrum of a $D_{6d}$ point group structure with $N=300$, where the Hamiltonian, irreducible representations and optical conductivity are obtained in Appendixes~\ref{sec:Hamiltonian}, ~\ref{sec:irreps} and ~\ref{sec:opcond}, respectively. The Fermi energy has been shifted to $0$ eV. We point out that the Fermi level for undoped TBGQDs is usually not at zero anymore because of the next-nearest neighbor and third-nearest-neighbor hoppings taken into account. Therefore, the Fermi energy should be firstly determined by the half filling rule such that the obtained optical conductivity is intrinsic. We can see a series of absorption peaks in Figs. \ref{fig:OPs_D6d}(b) and \ref{fig:OPs_D6d}(c) from infrared to ultraviolet frequency, such as peak $1$: $E_1 \rightarrow E_2$ at $\hbar\omega=0.255t_0$, peak $2$: $E_5 \rightarrow E_4$ at $\hbar\omega=0.33t_0$, and peak $3$: $A_1 \rightarrow E_1$ at $\hbar\omega=0.485t_0$. In addition, the optical conductivity is obviously isotropic, i.e., $\sigma_{xx} = \sigma_{yy}$. Although the band gap with $E_g=0.2t_0$ is determined by the energy difference between the highest occupied $E_1$ state and the lowest unoccupied state $E_4$ state in Fig. \ref{fig:OPs_D6d}(a), the transition $E_1 \leftrightarrow E_4$ is forbidden. These absorption peaks and forbidden transitions are a result of the selection rules of $D_{6d}$ point group as shown in Table~\ref{table:sel}. The optical conductivity spectrum for other $9$ point group structures are also calculated (see supplemental Figs. S1-S3\cite{SM}) with corresponding absorption peaks following the selection rules as well.

\begin{figure*}[!htbp]
\centering
\includegraphics[width=17 cm]{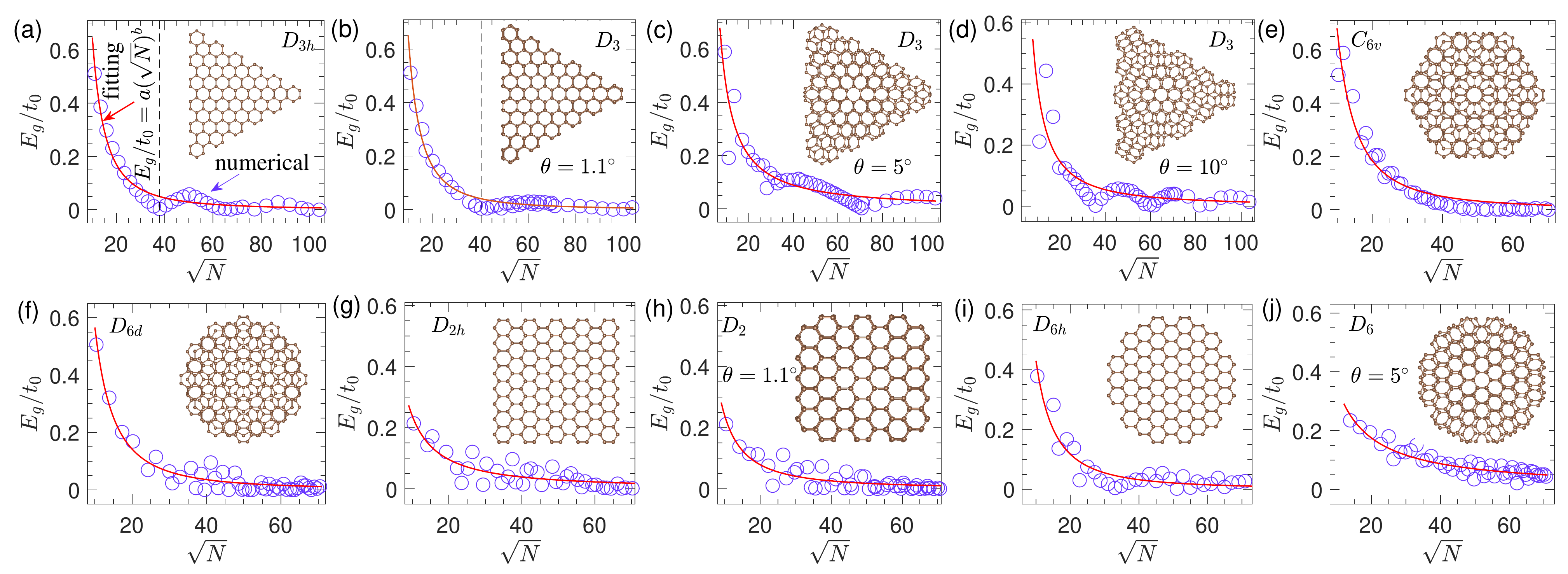}
\caption{{Band gap scaling law of TBGQDs.} In each subplot from (a) to (j), each inset as an example shows the structure of quantum dot with its point group. The numerical results of $E_g/t_0$ as a function of $\sqrt{N}$ are denoted by the blue circles, and the power-law fitting of $E_g/t_0 =a(\sqrt{N})^b$ is plotted by the red line. Here, $U=0$, $N$ is the number of C atoms, and the fitted values of $a$ and $b$ are listed in in Table~\ref{table:scalingab}.}
\label{fig:bandgapsclaing}
\end{figure*}

\begin{table*}
\caption{{Fitted values of dimensionless $a$ and $b$ for $U=0$, $U=0.6t_0$ and $U=1.2t_0$.} These considered TBGQDs with their point groups and twist angles are the same as those in Fig. \ref{fig:bandgapsclaing}. }
\label{table:scalingab}
\begin{ruledtabular}
\begin{centering}
\begin{tabular}{cccccccc}
& & \multicolumn{2}{c}{$U=0$} & \multicolumn{2}{c}{$U=0.6t_0$}  & \multicolumn{2}{c}{$U=1.2t_0$} \\
\cline{3-4} \cline{5-6} \cline{7-8}
Point groups & Twist angle $(\theta)$ &$a$ & $b$ & $a$ & $b$ & $a$ & $b$ \\
\hline
$D_{3h}$ & $0^{\circ}$ &$54.889$ &$-1.942$ &$54.403$ &$-1.926$ &$53.652$ &$-1.920$ \\
$D_{3}$ & $1.1^{\circ}$ &$62.217$ &$-1.980$ &$61.911$ &$-1.978$ &$61.477$ &$-1.975$ \\
$D_{3}$ & $5^{\circ}$ &$14.894$ &$-1.395$ &$15.410$ &$-1.408$ &$15.651$ &$-1.413$ \\
$D_{3}$ & $10^{\circ}$ &$95.092$ &$-2.103$ &$96.872$ &$-2.110$ &$95.553$ &$-2.104$ \\
$C_{6v}$ & $30^{\circ}$ &$52.096$ &$-1.884$ &$37.178$ &$-1.749$ &$19.987$ &$-1.501$ \\
$D_{6d}$ & $30^{\circ}$ &$55.373$ &$-1.991$ &$49.765$ &$-1.947$ &$23.831$ &$-1.645$ \\
$D_{2h}$ & $0^{\circ}$ &$4.832$ &$-1.307$ &$5.084$ &$-1.329$ &$7.142$ &$-1.464$ \\
$D_{2}$ & $1.1^{\circ}$ &$9.399$ &$-1.595$ &$7.805$ &$-1.521$ &$10.513$ &$-1.644$ \\
$D_{6h}$ & $0^{\circ}$ &$32.637$ &$-1.879$ &$32.640$ &$-1.879$ &$30.506$ &$-1.851$ \\
$D_{6}$ & $5^{\circ}$ &$3.500$ &$-0.995$ &$3.430$ &$-0.988$ &$3.429$ &$-0.988$ \\
\end{tabular}
\par\end{centering}
\end{ruledtabular}
\end{table*}

To further understand how the allowed transitions for both linearly and circularly polarized light are identified by symmetry operators $\hat{O}$ and $\hat{C}_n$, respectively, we first correspondingly classify the energy levels via $\gamma_l$ of $\hat{O}$ and $\phi_l$ of $\hat{C}_n$, as shown in Fig. \ref{fig:sel_D6d}, where $\hat{O}=\hat{\sigma}_d(xz)$ with reflection plane $xz$ and $\hat{C}_n = \hat{C}_6(z)$ for the structure with $N=300$ and $D_{6d}$ point group in Fig. \ref{fig:OPs_D6d}. As we can see, under the reflection operator $\hat{\sigma}_d$ in Fig. \ref{fig:sel_D6d}(b), the original energy levels in Fig. \ref{fig:sel_D6d}(c) are divided into two columns of energy levels denoted by $\gamma_l=1$ and $\gamma_l=-1$ in Fig. \ref{fig:sel_D6d}(a). Under the rotational operator $\hat{C}_6(z)$ in Fig. \ref{fig:sel_D6d}(d), the original energy levels in Fig. \ref{fig:sel_D6d}(c) are separated into $6$ columns of energy levels denoted by $\phi_l=0, 1, 2, 3, 4, 5$ in Fig. \ref{fig:sel_D6d}(e).
For linearly polarized light, $\hat{j}_x$ and $\hat{j}_y$ under the transformation of $\hat{\sigma}_d$ satisfy $\hat{\sigma}^\dagger_d\hat{j}_x\hat{\sigma}_d = \hat{j}_x$ and $\hat{\sigma}^\dagger_d\hat{j}_y\hat{\sigma}_d = -\hat{j}_y$. It means that $q_x=1$ and $q_y=-1$. Therefore, the selection rules in Eq.~\eqref{eq:lselrule} require that $\gamma^*_k \gamma_l - 1 =0$ and $\gamma^*_k \gamma_l +1  =0$ for $x$ and $y$ linearly polarized light, respectively. Consequently, the calculated matrix elements of current operators $\braket {\psi_k |\hat{j}_x |\psi_l}$ and $\braket {\psi_k |\hat{j}_y |\psi_l}$ display off-diagonal and diagonal patterns within the classified states of $\hat{\sigma}_d$, as demonstrated in Figs. \ref{fig:sel_D6d}(f) and \ref{fig:sel_D6d}(g). Similarly, for right and left circularly polarized light, $\braket {\psi_k |\hat{j}_+ |\psi_l}$ and $\braket {\psi_k |\hat{j}_-|\psi_l}$ within the classified states of $\hat{C}_6(z)$ correspondingly obey $\Delta \phi =1$ and $\Delta \phi =-1$ in Eq.~\eqref{eq:cselrule}, as demonstrated in Figs. \ref{fig:sel_D6d}(h) and \ref{fig:sel_D6d}(i). Since $D_2$ and $D_{2h}$ have intrinsically anisotropic properties, we calculate these matrix elements of current operators for structures with other $7$ point group structures and find similar results (see supplemental Figs. S4-S10\cite{SM}) governed by the polarization-dependent selection rules in Eq.~\eqref{eq:cselrule} and Eq.~\eqref{eq:lselrule}.

\begin{figure*}[!htbp]
\centering
\includegraphics[width=17 cm]{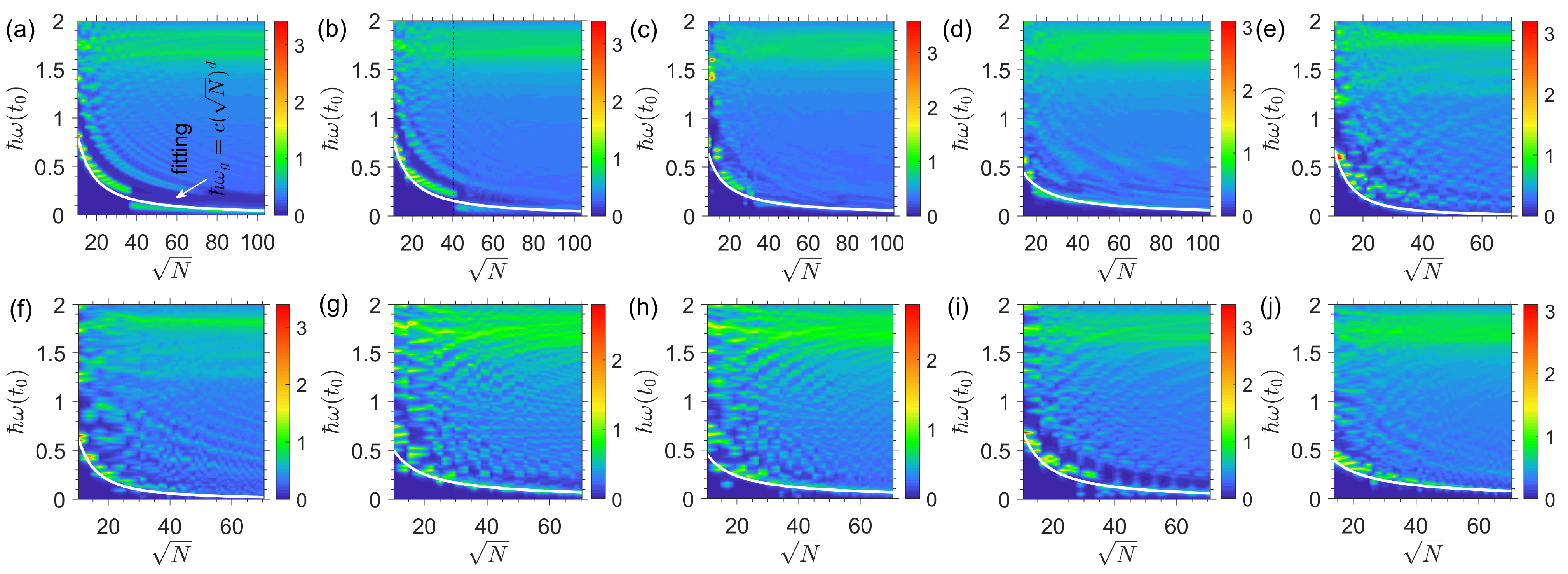}
\caption{{Optical conductivity contour plots for $U=0$.} In each subplot from (a) to (j), the real part of optical conductivity $\sigma_{xx}$ in unit of $\sigma_0 = \pi e^2/(4\hbar)$ is plotted as a function of $\sqrt{N}$ and $\hbar\omega$, with the same structure and point group as those for each $N$ in each subplot of Fig. \ref{fig:bandgapsclaing}. The white line represents the power-law fitting of $\hbar\omega_g/t_0 =c(\sqrt{N})^d$, with $\omega_g$ as the frequency of the first main absorption peak. }
\label{fig:opgapscaling}
\end{figure*}

\section{Band gap scaling}

Besides the theoretical interest itself, the size dependence of the band gap is critical to practical designing and engineering of nanoelectronics. As shown in method sections, the size (such as the edge length or radius) of TBGQDs is proportional to $\sqrt{N}$, and hence it is reasonable to use $\sqrt{N}$ as the size index. Figs. \ref{fig:bandgapsclaing}(a)-\ref{fig:bandgapsclaing}(j) present the band gap of TBGQDs with $U=0$ as a function of $\sqrt{N}$, for $D_{3h}$, $D_3 (\theta = 1.1^\circ)$, $D_3 (\theta = 5^\circ)$, $D_3 (\theta = 10^\circ)$, $C_{6v}$, $D_{6d}$, $D_{2h}$, $D_2 (\theta = 1.1^\circ)$, $D_{6h}$ and $D_6 (\theta = 5^\circ)$ point groups, respectively. We can see that, the band gap exhibits firstly a rapid decay with respect to $\sqrt{N}$ from few hundreds to several thousands of $N$, and then converges with an oscillation towards zero because TBGQDs recover the electronic spectrum of twisted bilayer graphene with zero gap in the limit of huge $N$. It is known that the band gap follows $1/R^2$ power law with a radius $R$ for usual semiconductor quantum dots\cite{chakraborty1999}. For graphene ribbons, the scaling law obeys approximately $1/W$ (i.e., $1/\sqrt{N}$) and $1/W^2$ (i.e., $1/N$) relations with ribbon width $W$ for zigzag and armchair edges, respectively, because of their corresponding linear and parabolic energy dispersions\cite{katsnelson2020graphene}. Therefore, a power-law of $E_g/t_0 =a(\sqrt{N})^b$ is adopted to fit the decay of the band gap of TBGQDs. These fitted values of dimensionless numbers $a$ and $b$ are listed in Table~\ref{table:scalingab}. For TBGQDs with different point groups or same point groups but with different twist angles, some of these structures display relatively close values of $a$ and $b$, and some exhibit different $a$ and $b$. However, all the values of the power index $b$ are almost inside $[-2, -1]$. The reason is that here TBGQDs are generally customized with a random edge profile instead of pure zigzag or armchair edges such that the band gap scaling behaves like a mixed behavior of relativistic and nonrelativistic particles.

\begin{table*}
\caption{{Fitted values of scaling indexes $c$ and $d$ for $U=0$, $U=0.6t_0$ and $U=1.2t_0$.} These considered TBGQDs with their point groups and twist angles are the same as those for each $N$ in Fig. \ref{fig:bandgapsclaing}.}
\label{table:scalingcd}
\begin{ruledtabular}
\begin{centering}
\begin{tabular}{cccccccc}
& & \multicolumn{2}{c}{$U=0$} & \multicolumn{2}{c}{$U=0.6t_0$}  & \multicolumn{2}{c}{$U=1.2t_0$} \\
\cline{3-4} \cline{5-6} \cline{7-8}
Point groups & Twist angle $(\theta)$ &$c$ & $d$ & $c$ & $d$ & $c$ & $d$ \\
\hline
$D_{3h}$ & $0^{\circ}$ &$16.477$ &$-1.272$ &$16.373$ &$-1.270$ &$16.373$ &$-1.270$ \\
$D_{3}$ & $1.1^{\circ}$ &$14.199$ &$-1.221$ &$14.529$ &$-1.229$ &$15.443$ &$-1.254$ \\
$D_{3}$ & $5^{\circ}$ &$14.699$ &$-1.258$ &$15.113$ &$-1.268$ &$15.063$ &$-1.267$ \\
$D_{3}$ & $10^{\circ}$ &$10.182$ &$-1.142$ &$10.630$ &$-1.158$ &$10.350$ &$-1.147$ \\
$C_{6v}$ & $30^{\circ}$ &$57.320$ &$-1.896$ &$23.456$ &$-1.545$ &$15.717$ &$-1.385$ \\
$D_{6d}$ & $30^{\circ}$ &$36.626$ &$-1.745$ &$25.501$ &$-1.599$ &$12.850$ &$-1.325$ \\
$D_{2h}(x)$ & $0^{\circ}$ &$5.459$ &$-1.022$ &$5.459$ &$-1.022$ &$5.353$ &$-1.015$ \\
$D_{2h}(y)$ & $0^{\circ}$ &$1.728$ &$-0.788$ &$1.728$ &$-0.788$ &$1.727$ &$-0.784$ \\
$D_{2}(x)$ & $1.1^{\circ}$ &$4.935$ &$-1.021$ &$4.516$ &$-0.988$ &$4.466$ &$-0.980$ \\
$D_{2}(y)$ & $1.1^{\circ}$ &$1.901$ &$-0.839$ &$1.825$ &$-0.820$ &$1.524$ &$-0.756$ \\
$D_{6h}$ & $0^{\circ}$ &$12.462$ &$-1.247$ &$12.462$ &$-1.247$ &$14.428$ &$-1.314$ \\
$D_{6}$ & $5^{\circ}$ &$5.385$ &$-0.994$ &$5.443$ &$-0.987$ &$5.443$ &$-0.987$ \\
\end{tabular}
\par\end{centering}
\end{ruledtabular}
\end{table*}

We further reveal the effects of local Coulomb interaction on the scaling of TBGQDs. The band gaps as a function of $\sqrt{N}$ for these TBGQDs with $U=1.2t_0$\cite{schuler2013optimal,yazyev2010emergence} and $U=0.6t_0$ are calculated and plotted in supplemental Fig. S11\cite{SM} and Fig. S12\cite{SM}, respectively. The latter can be viewed as an effective dielectric screening for the former with a dielectric constant $2$ from external environment if a substrate is used. As a consequence of Coulomb interactions, the edge magnetism in a few structures possibly emerges even with random edge profiles, and hence the ground states are determined by minimizing the total energy after the comparison among the ferromagnetic, antiferromagnetic and non-magnetic states. After the power-law fitting we obtain the dimensionless numbers $a$ and $b$ and also list them in Table~\ref{table:scalingab} for comparison with the case under $U=0$. As we can see, the local Coulomb interaction $U$ has weak influence on $a$ and $b$ for a structure with fixed point group. It means that the effective tight-binding kinetic energy term of Hamiltonian in Eq.~\eqref{eq:hamiltonian} can approximately capture the main scaling behaviors. Compared with the weak influence of the local Coulomb interaction, the twist angle has a remarkable impact on the size scaling of band gap. Here we also note that, (i) quantum dots with odd number of C atoms such as $C_{3v}$ structures are excluded because of its zero band gap according to half-filling, and (ii) small structures with several tens of C atoms are excluded because of more strong confinement effect, where configuration interaction method\cite{gucclu2014graphene} and quantum Monte Carlo simulation\cite{feldner2010magnetism} are alternative methods to describe the correlation phenomenon in these systems.

\begin{figure*}[!htbp]
\centering
\includegraphics[width=16 cm]{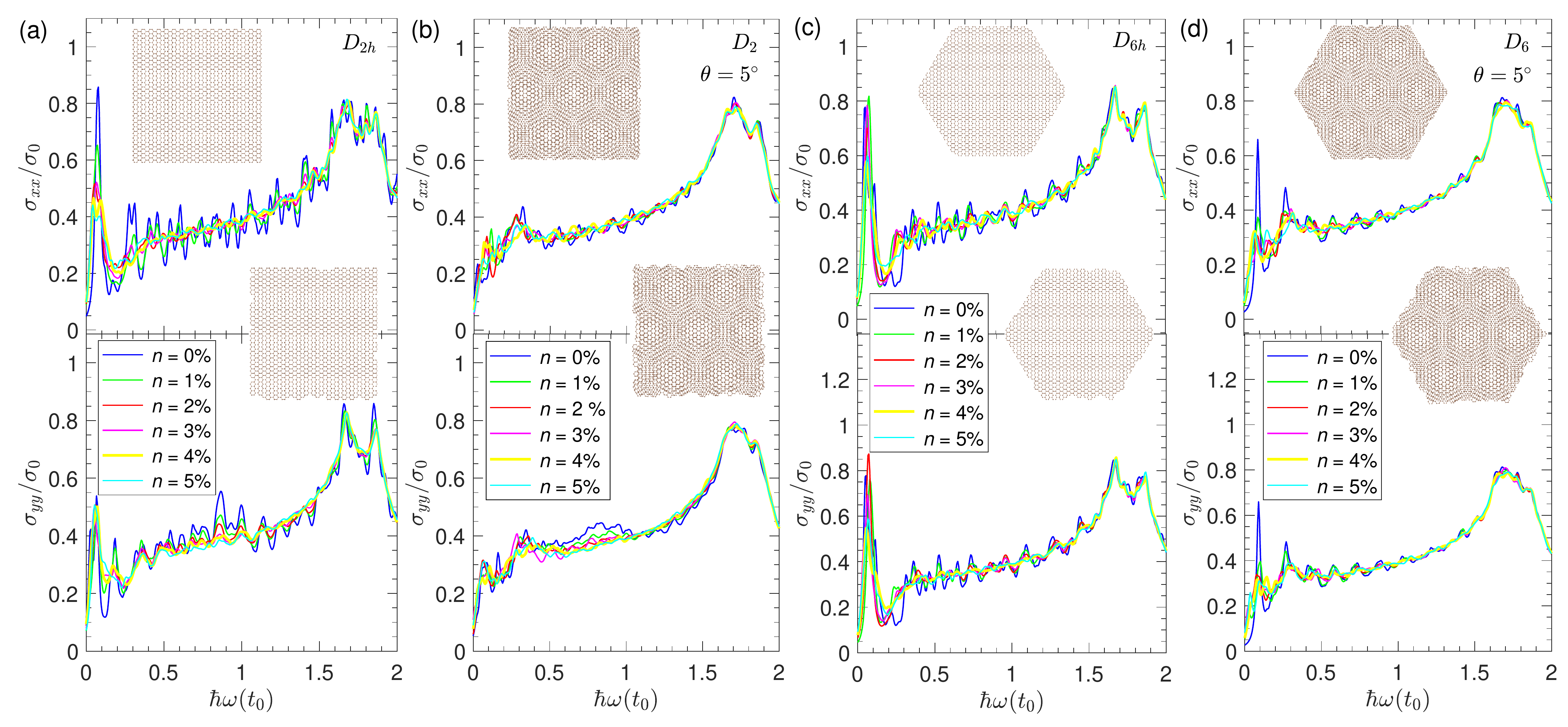}
\caption{{Edge atom vacancy effects on optical conductivity spectrum.} The real parts of optical conductivity $\sigma_{xx}$ (up panel) and $\sigma_{yy}$ (down panel) in unit of $\sigma_0 = \pi e^2/(4\hbar)$ are plotted as a function of $\hbar\omega$ with different vacancy densities $n$ for $D_{2h}$ in (a), $D_{2}$ in (b), $D_{6h}$ in (c) and $D_{6h}$ in (d). The up (down) insets show the the quantum dot structures without (with) random edge atom vacancies, where the number of atoms are $N=4148$, $N=4264$, $N=4212$ and $N=4344$, respectively, for these structures without vacancies in the up insets. }
\label{fig:vacancy}
\end{figure*}

\section{Optical band gap scaling}

Following the selection rules in Table~\ref{table:sel}, the optical conductivity spectrum of these structures in Fig. \ref{fig:bandgapsclaing} as a function of size indice $\sqrt{N}$ and photon energy $\hbar\omega$ exhibit two remarkable absorption characteristics, as shown in Fig. \ref{fig:opgapscaling}. Firstly, a relatively strong absorption region exists at about $1.5t_0<\hbar\omega<2t_0$ and changes little with the size variation for all structures. Such a stable absorption region actually corresponds to these interband transitions between these energy levels near the Van Hove singularities of the infinite size twisted bilayer graphene (see supplemental Fig. S13\cite{SM}). Secondly, a forbidden absorption region with blue color below the white  fitted line diminishes with the increasing of the size. It means that, following the decay of the electronic band gap, the optical band gap $(\hbar\omega_g)$ also decreases with increasing the size. However, the fitting results in Table~\ref{table:scalingcd} from the power-law of $\hbar\omega_g/t_0 =c(\sqrt{N})^d$ show that the dimensionless numbers $c$ and $d$ are different from $a$ and $b$ of the electronic band gap. The reason is that, the interband transitions between the highest occupied and lowest unoccupied energy levels are only allowed if their transitions obey the optical selection rules, otherwise, the transitions are forbidden.
In addition, two other aspects should also be figured out: (i) $D_{2h}$ and $D_2$ point group structures exhibit an anisotropic scaling behavior (see $\sigma_{yy}$ in supplemental Fig. S14\cite{SM}), and (ii) $D_{3h}$ and $D_3$ point group structures in Figs. \ref{fig:opgapscaling}(a) and \ref{fig:opgapscaling}(b) display a piecewise decay behavior because of the oscillating decay of electronic structures as shown in Figs. \ref{fig:bandgapsclaing}(a) and \ref{fig:bandgapsclaing}(b) with zero band gap at $N = 1440$ and $N = 1632$ denoted by the vertical dash lines, respectively. The local electron-electron interaction has weak influence on both the optical conductivity spectrum and optical band gap as shown in supplemental Fig. S15\cite{SM} and Fig. S16\cite{SM} and Table~\ref{table:scalingcd}.

\section{Structure relaxations, edge atom vacancies and twist angle effects on optical spectrum}

We now reveal the effects of structure relaxations (see Appendix~\ref{sec:str}) and edge atom vacancies on the optical conductivity spectrum of TBGQDs. As seen in supplemental Fig. S17\cite{SM}, the structure relaxation has weak modifications on the frequency and magnitude of absorption peaks for these structures. As an example, Fig. \ref{fig:vacancy} shows the optical conductivity spectrum of quantum dots for $D_{2h}$, $D_2$, $C_{6v}$ and $D_{6d}$ point groups with different vacancy defect density $n$. With the increase of $n$, the disorder is enhanced, and the original point group symmetry is also broken by the nonuniform edge atom vacancies. It can be seen that, these absorption peak magnitudes are weakened by the increased $n$, but the edge vacancy has weak influences on the frequency positions of these main absorption peaks and the strong absorption region associated with the van Hove singularity. It means that in these structures with edge vacancy defects and slightly broken point group symmetry the main optical absorption features remain.

\begin{figure}[!htbp]
\centering
\includegraphics[width=7.5 cm]{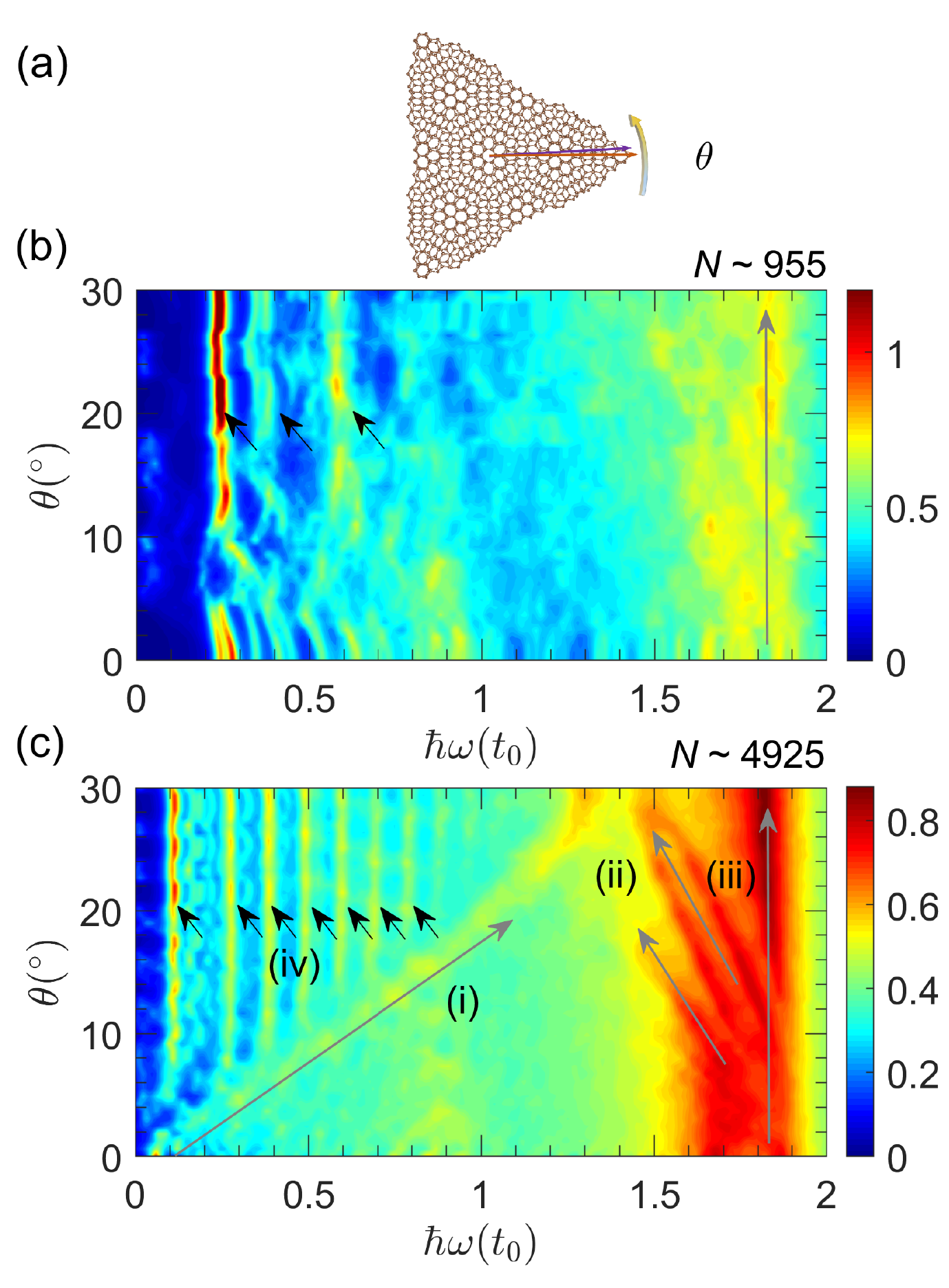}
\caption{{Twist angle dependence of optical spectrum.} (a) A schematic structure of a triangular quantum dot with the hexagon center and a fixed $x$ axis along the armchair direction before rotation. The real parts of optical conductivities $\sigma_{xx}$ in unit of $\sigma_0 = \pi e^2/(4\hbar)$ are plotted as a function of the twist angle $\theta$ and the photon energy $\hbar\omega$ in (b) with $N\sim 955$ and in (c) with $N\sim 4925$. }
\label{fig:angle}
\end{figure}

We further evaluate the twist angle dependence of optical absorption spectrum of quantum dots with slightly variable $N$, which is inevitably induced in experiments because of the edge atom vacancies. We first consider a typically triangular quantum dot with the hexagon center and $x$ axis along armchair direction. The optical conductivities of the structures with $\sqrt{S}\sim 5$ nm $(N\sim 955)$ and $\sqrt{S}\sim 11.4$ nm $(N\sim 4925)$ as a function of the twist angle and photon energy $\hbar\omega$ are shown in Fig. \ref{fig:angle}. For a relatively small structure with $N \sim 955$, beside the absorption region associated with the van Hove singularity, the size-induced three absorption peaks at $0.2t_0<\hbar\omega<0.6t_0$ from infrared to visible frequancy change little when $\theta > 10^\circ$, while there are multiple peaks when $\theta$ is below $10^\circ$, as shown in Fig. \ref{fig:angle}(b). For a relatively large structure with $N \sim 4925$, we can observe four groups of characteristic conductivity peaks in Fig. \ref{fig:angle}(c). With the increase of twist angle, group (i) peaks move toward high frequency, group (ii) peaks move toward low frequency, and group (iii) peaks associated with the van Hove singularity change little. The three characteristics are similar to those of infinite size twisted bilayer graphene\cite{moon2013optical,le2018electronic,vela2018electronic}. However, besides the former three group peaks, there exists the fourth group (iv) peaks with multiple discrete absorption frequency from infrared to ultraviolet light, and they are almost independent of the twist angle above the angles of peaks (i). These peaks are missing in 2D infinite size twist systems and they arise from the multiple interband transitions between these discrete energy levels induced by the quantum confinement effect of finite size. In addition, group (iv) peaks also exist in other structures (see hexagonal quantum dot in supplemental Fig. S18\cite{SM}). Therefore, these size-induced optical absorptions from infrared to visible and ultraviolet frequencies will enable TBGQDs to have important applications on photovoltaic devices and photodetectors. At last, we also figure out that the temperature has weak influences on the optical conductivity of TBGQDs as shown in supplemental Fig. S13\cite{SM}.

\section{Conclusion}

In conclusion, we construct the polarization-dependent selection rules for systems with point group symmetry. The rotational quantum number change characterizes the selection rules of circularly polarized light. The eigenvalues of symmetry operator $\hat{O}$ (such as $2$-fold rotational operator or reflection operator) characterize the selection rules of linearly polarized light in $D_{n}$, $D_{nh}$, $D_{nd}$ and $C_{nv}$ systems. We have designed and classified various TBGQDs into $10$ different point group structures which are feasible in currently experimental fabrications. The optical selection rule databases for all of these quantum dots are made. The calculated current operator matrix elements identify our polarization-dependent selection rules. The band gap scaling of TBGQDs follows a power-law with a power index inside $[-2,-1]$, and the twist degree of freedom has a remarkable impact on the size scaling compared with the weak influences of local Coulomb interactions. We made an atlas of both size-dependent and twist-angle-dependent optical spectra of TBGQDs. The optical band gap scaling also follows a power-law but with its power index less than that of electronic band gap as a result of selection rules. Besides the three groups of optical conductivity peaks appearing in infinite twisted bilayer graphene, another new group of peaks with multiple discrete absorption frequencies from infrared to ultraviolet light occurs in quantum dot systems because of the quantum confinement effect of finite size. These new peaks render TBGQDs for applications on photovoltaic devices and photodetectors. The mapped atlas and constructed selection rule database of optical spectrum present a comprehensive structure/symmetry-function interrelation and allows an excellent geometrical control of optical properties for TBGQDs as a building block in on-chip carbon optoelectronics\cite{areshkin2007,payod20202n}.

\begin{acknowledgments}
S.Y. acknowledges the support by the National Science Foundation of China (No. 11774269). H.-Q. L. acknowledges the financial support from NSAF (No. U1930402) and NSFC (No. 11734002). G.Y. and Y.W. acknowledge the support from China Postdoctoral Science Foundation (Grant Nos. 2018M632902 and 2019M660433) and NSFC (No. 11832019). MIK acknowledges the support by the ERC Synergy Grant, project 854843 FASTCORR.

Y.W. and G.Y. contributed equally to this work.
\end{acknowledgments}

\begin{appendix}

\section{Procedures for generating TBGQDs} \label{sec:pgQDs}

The process involves two steps to generate TBGQDs. Firstly, we define the lattice vectors for each layer and the rotation center between the two layers such that we can construct an infinite twisted bilayer graphene, where the atom sites are described by
\begin{equation}
\begin{split}
\bm{R}_{l,A(B)} &=  m_l\bm{a}_{l,1}+n_l\bm{a}_{l,2} + \bm{\tau}_{l,A(B)}. \\
\end{split}
\end{equation}
The layer index $l=1,2$ denotes layer $1$ and layer $2$, respectively, $\bm{a}_{l,1}$ and $\bm{a}_{l,2}$ are the basis vectors of each layer, $m_l$ and $n_l$ are arbitrary integers, and $\bm{\tau}_{l,A(B)}$ is the relative vectors of sublattices $A$ and $B$ inside a unit cell for each layer. For monolayer graphene before twist, if the armchair edge is along the $x$ axis, the basis vectors read
\begin{equation}
\begin{split}
\bm{a}_{1} = \frac{\sqrt{3}a}{2}\hat{x}-\frac{a}{2}\hat{y},\ \bm{a}_{2} = \frac{\sqrt{3}a}{2}\hat{x}+\frac{a}{2}\hat{y}.
\label{eq:arm}
\end{split}
\end{equation}
If the zigzag edge is along the $x$ axis, the basis vectors read
\begin{equation}
\begin{split}
\bm{a}_{1} = a\hat{x}+0\hat{y},\ \bm{a}_{2} = \frac{a}{2}\hat{x}+\frac{\sqrt{3}a}{2}\hat{y}.
\label{eq:zig}
\end{split}
\end{equation}
After a twist between the two monolayers with angle $\theta$, we can write the basis vectors of layers $1$ and $2$ as $\bm{a}_{1,i}=\bm{R}_{-\theta/2}\cdot \bm{a}_{i}$ and $\bm{a}_{2,i}=\bm{R}_{\theta/2}\cdot \bm{a}_{i}$ with $\bm{R}_\theta$ as the rotation operation. The rotation geometrical centers can be at atom, hexagon center and bond center, where their relative vectors of sublattices are respectively given by
\begin{equation}
\bm{\tau}_{l,A} = \bm{0},\ \bm{\tau}_{l,B} = \frac{1}{3}\bm{a}_{l,1} + \frac{1}{3}\bm{a}_{l,2},
\label{eq:tau_atom}
\end{equation}
\begin{equation}
\bm{\tau}_{l,A} = \frac{1}{3}\bm{a}_{l,1} + \frac{1}{3}\bm{a}_{l,2},\ \bm{\tau}_{l,B} = \frac{2}{3}\bm{a}_{l,1} + \frac{2}{3}\bm{a}_{l,2},
\label{eq:tau_hex}
\end{equation}
\begin{equation}
\bm{\tau}_{l,A} = -\frac{1}{6}\bm{a}_{l,1} - \frac{1}{6}\bm{a}_{l,2},\ \bm{\tau}_{l,B} = \frac{1}{6}\bm{a}_{l,1} + \frac{1}{6}\bm{a}_{l,2}.
\label{eq:tau_bond}
\end{equation}
In the second step, we cut the infinite twisted bilayer graphene into nanoflakes by leaving the atoms only inside a desired polygon and remove the edge atoms with two dangling bonds to reduce defect states. We now show in details how to chose the rotation geometrical centers and relative vectors of sublattices $\bm{\tau}_{l,A(B)}$ to generate TBGQDs with $10$ different point group symmetries, respectively.

For structures with point groups $D_n$ and $D_{nh}$ with $n=2,3,6$, the twist angles $\theta=0$ and $\theta\neq0$ are respectively required for $D_{nh}$ and $D_{n}$. For $D_{2h}$ and $D_{2}$, we need to cut the infinite bilayers into rectangle nanoflakes with two vertical sides and two horizontal sides. As an example, we have chose the bond center as the geometrical center with $\bm{\tau}_{l,A(B)}$ in Eq.~\eqref{eq:tau_bond} for quantum dots with $D_{2h}$ and $D_{2}$ as shown in Figs. \ref{fig:structures}a and \ref{fig:structures}b. For $D_{3h}(D_{6h})$ and $D_{3}(D_{6})$, we need to cut the infinite bilayers into triangular (hexagonal) nanoflakes with three (six) sides. The hexagon center is chosen as the geometrical center in quantum dots with $D_{3h}$ in Fig. \ref{fig:structures}c, $D_{3}$ in Fig. \ref{fig:structures}d, $D_{6h}$ in Fig. \ref{fig:structures}g and $D_{6}$ as shown in Fig. \ref{fig:structures}h.

For structures with point group $D_{3d}$, the geometrical center is at the C atom with its $\bm{\tau}_{l,A(B)}$ in Eq.~\eqref{eq:tau_atom}, and the twist angle is zero. The armchair edge is chosen along the $x$ axis with its the basis vectors in Eq.~\eqref{eq:arm}. The infinite bilayers are then cut into triangular nanoflakes as shown in Fig. \ref{fig:structures}f.

For structures with point groups $C_{3v}$, $C_{6v}$ and $D_{6d}$, the geometrical center is always at the hexagon center with its $\bm{\tau}_{l,A(B)}$ in Eq.~\eqref{eq:tau_hex}. The bottom and top layers have armchair and zigzag edge along the $x$ axis with their basis vectors in Eqs.~\eqref{eq:arm} and \eqref{eq:zig}, respectively. It also means that the twist angle $\theta$ is $30^{\circ}$. Then, the infinite bilayers are cut into triangular, hexagonal and dodecagonal nanoflakes for $C_{3v}$, $C_{6v}$, and $D_{6d}$, respectively.

\section{Hamiltonian} \label{sec:Hamiltonian}

For twisted bilayer graphene systems, the Hubbard Hamiltonian including both the effective tight-binding kinetic energy contributed by the $p_z$ orbital of C atoms and the electron-electron interactions reads,
\begin{equation}
H=\sum_{i,s}\varepsilon_in_{i,s}+\sum_{i,\mathbf{r}_{ij},s}t(\mathbf{r}_{ij})
c^\dag_{i,s}c_{j,s}+U\sum_in_{i,\uparrow}n_{i,\downarrow},
\label{eq:hamiltonian}
\end{equation}
where $n_{i,s}=c^\dag_{i,s}c_{i,s}$ with the spin index $s$, $\varepsilon_i$ is the on-site energy and has been set as zero, $U$ is an effective on-site Coulomb repulsion (i.e., $U^*$\cite{schuler2013optimal}, here labeled as $U$ for simplicity), and the hopping $t(\mathbf{r}_{ij})$ is written as a function of $\mathbf{r}_{ij}$, i.e.,
\begin{equation}
t(\mathbf{r}_{ij})=V_{pp\sigma}(|\mathbf{r}_{ij}|)cos^2\beta+V_{pp\pi}(|\mathbf{r}_{ij}|)sin^2\beta,
\label{tb:r}
\end{equation}
with $\beta=\hat{z}\cdot \mathbf{r}_{ij}/|\mathbf{r}_{ij}|$. The Slater-Koster bond integrals take the forms as
\begin{equation}
\begin{split}
V_{pp\sigma}(|\mathbf{r}_{ij}|)&=-t_0e^{2.218(b_0-|\mathbf{r}_{ij}|)}F(|\mathbf{r}_{ij}|),\\
V_{pp\pi}(|\mathbf{r}_{ij}|)&=t_1e^{2.218(h-|\mathbf{r}_{ij}|)}F(|\mathbf{r}_{ij}|),\\
\end{split}
\label{tb:SK}
\end{equation}
where $t_0=2.8$ eV, $b_0=1.42$ {\AA}, $t_1 = 0.48$ eV, $F(|\mathbf{r}_{ij}|)=1/(1+e^{(|\mathbf{r}_{ij}|-0.265)/5})$, and $h$ is the interlayer distance with $3.35$ {\AA}. In our calculations, the cutoff carbon-carbon hopping distance is $5$ {\AA}. The $p_z$-orbital based tight-binding
model in Eq.~\eqref{tb:r} and Eq.~\eqref{tb:SK} has been widely used to well describe the electronic structures in twisted bilayer graphene systems\cite{trambly2010localization,ahn2018dirac,huder2018electronic,yu2019dodecagonal,kerelsky2019maximized,shi2020large,yu2021interlayer}.
The low-energy physics from the $p_z$-orbital based tight-binding model is also consistent with that from Wannier-function based tight-binding method\cite{fang2016electronic} and density functional theory\cite{yu2021interlayer}.
For the correlated $sp^2$ carbon systems with a moderate local $U$, the mean-field approximation can successfully capture the low-energy physics occurring in the systems\cite{fernandez2007magnetism,gucclu2014graphene,yazyev2010emergence,feldner2010magnetism}. Within this approximation, the two-body interactions can be decoupled as $n_{i,\uparrow}n_{i,\downarrow}\approx n_{i,\uparrow}\langle n_{i,\downarrow}\rangle+\langle n_{i,\uparrow} \rangle n_{i,\downarrow}-\langle n_{i,\uparrow} \rangle \langle n_{i,\downarrow}\rangle$. With the help of a self-consistent iterative calculation with a high convergence precision of $10^{-6}$ of $n_{i,s}$, the electronic structure and property of system are determined after minimizing the total energy.

\section{Irreducible representations} \label{sec:irreps}

For an eigen state with its energy $E_n$ and wave function $\psi^\xi_n$ (where $\xi=1,\cdots,g$, with the degree of degeneracy $g$), the representation matrix element $D^n_{\xi,\xi^\prime}(\hat{R})$ for an operation class $\hat{R}$ is written as $D^n_{\xi,\xi^\prime}(\hat{R})=\langle \psi^\xi_n |\hat{R}| \psi^{\xi^\prime}_n\rangle$. The character $\chi^n(\hat{R})$ of the representation of $\hat{R}$ is determined by Tr$[D^n(\hat{R})]$. We further obtain the reducible representation $\Gamma_{E_n}=\sum^\oplus_\mu a_\mu\Gamma_\mu$ of the energy state $E_n$  by virtue of the number of times $\Gamma_\mu$, i.e.,
\begin{equation}
a_\mu=\frac{1}{h}\sum_i \chi^n(\hat{R}_i)[\chi^n_{\Gamma_\mu}(\hat{R}_i)]^*,
\end{equation}
where $h$ is the order of point group, and $\hat{R}_i$ is an arbitrary symmetry operation of system.

\section{Optical conductivity} \label{sec:opcond}

According to the Kubo-Greenwood formula, the real part of interband optical conductivity (Re$[\sigma_{\alpha\alpha}]$,  neglecting the notation Re hereinafter for simpilicity) for each spin reads
\begin{equation}
\sigma_{\alpha\alpha}(\omega)=\frac{\pi\hbar}{A}\sum_{mn}\frac{f_{m}-f_{n}}{E_{n}-E_{m}}\frac{|\langle n|\hat{j}_\alpha|m\rangle|^2(\eta/\pi)}{(E_{m}-E_{n}-\hbar\omega)^2+\eta^2},
\label{eq:Kubo}
\end{equation}
where $A=NA_0$ is the area of quantum dot with $A_0=3\sqrt{3}b_0^2/4$ as the average area of each atom, $f_m(f_n)$ is the occupation number (Fermi-Dirac distribution function), the small smearing parameter $\eta$ is taken as $0.04$ eV in our calculations. The current operator $\hat{j}_\alpha$ is written as
\begin{equation}
\hat{j}_\alpha=e(\frac{-i}{\hbar})\sum_{ij}\sum_s(\mathbf{r}_{i,\alpha}-\mathbf{r}_{j,\alpha})
t(\mathbf{r}_{ij}) c^\dag_{i,s}c_{j,s},
\end{equation}
where $s$ is the spin index. Note that here in our calculations we have used $\sigma_0 = \pi e^2/(4\hbar)$ as the conductivity unit. On the other hand, the second factor inside the sum term in Eq.~\eqref{eq:Kubo} is actually a Dirac-delta operator function, which can be expressed as a Fourier transform of time evolution operator. In this respect, the real part of optical conductivity can also be written as\cite{yuan2010modeling}
\begin{equation}
\begin{split}
\sigma_{\alpha\alpha}(\omega)=\lim_{\eta\rightarrow 0^+}\frac{e^{-\hbar\omega/k_BT}-1}{\hbar\omega A}\int_0^\infty e^{-\eta\tau}sin\omega\tau \\
\times 2Im\langle\varphi_{2}(\tau)|\hat{j}_\alpha|\varphi_{1}(\tau)\rangle_\alpha,
\end{split}
\label{eq:Kubotime}
\end{equation}
where the wave functions $|\varphi_{1}(\tau)\rangle_\alpha$ and $|\varphi_{2}(\tau)\rangle$ take the forms as
\begin{equation}
\begin{split}
|\varphi_{1}(\tau)\rangle_\alpha &= e^{-iH\tau/\hbar}[1-f(H)]\hat{j}_\alpha|\varphi_0\rangle, \\
|\varphi_{2}(\tau)\rangle &= e^{-iH\tau/\hbar}f(H)|\varphi_0\rangle.
\end{split}
\label{eq:Kubotimewave}
\end{equation}
Here, $f(H)=1/(1+e^{(H-\mu)/k_BT})$ is Fermi-Dirac distribution operator with the chemical potential $\mu$, and $|\varphi_0\rangle$ is an initial state consisting of a random superposition of the $p_z$ orbitals at all sites. The the time-evolution based calculation in Eq.~\eqref{eq:Kubotime} is very effective for predicting the optical conductivity of systems with more than tens of thousands of atoms\cite{yuan2010modeling}, and hence here Eq.~\eqref{eq:Kubotime} is used for periodical systems with a lot of atoms (see supplemental Fig .S13\cite{SM}) for comparison with quantum dots.

\section{Structure relaxation} \label{sec:str}

The atomistic model based on the classical REBO\cite{rebo} (intra-layer interaction) and Kolmogorov/Crespi/full\cite{KCfull} (inter-layer interaction) potentials is implemented in LAMMPS software\cite{lammps0,lammps1}. All of the edge carbon atoms are saturated by hydrogen atoms for the relaxation.

\end{appendix}

\bibliography{References}

\end{document}